\documentclass{aa}  

\usepackage{graphicx}
\usepackage{txfonts}

\usepackage{natbib}
\bibpunct{(}{)}{;}{a}{}{,}

\usepackage{color}
\definecolor{mksc}{rgb}{0.0,0.8,0.2}

\begin{document} 

\title{MUSE tells the story of NGC 4371: The dawning of secular evolution}
\titlerunning{The dawning of secular evolution}
\authorrunning{Gadotti et al.}
%\subtitle{One example}

\author{Dimitri A. Gadotti
          \inst{1}
\and   Marja K. Seidel
          \inst{2,3}
\and   Patricia S\'anchez-Bl\'azquez
          \inst{4}
\and   Jesus Falc\'on-Barroso
          \inst{2,3}
\and   Bernd Husemann
          \inst{5}
\and   Paula Coelho
          \inst{6}
\and  Isabel P\'erez
          \inst{7,8}}

\institute{European Southern Observatory, Casilla 19001, Santiago 19, Chile\\
\email{dgadotti@eso.org}
\and   Instituto de Astrof\'isica de Can\'arias, E-38200 La Laguna, Tenerife, Spain
\and   Departamento de Astrof\'isica, Universidad de La Laguna (ULL), E-38206 La Laguna,
Tenerife, Spain
\and   Departamento de F\'isica Te\'orica, Universidad Aut\'onoma de Madrid, E-28049 Cantoblanco, Spain
\and   European Southern Observatory, Karl-Schwarzschild-Str. 2, 85748, Garching b. Muenchen, Germany
\and   Instituto Astron\^omico e Geof\'isico, Universidade de S\~ao Paulo, Rua do Mat\~ao, 1226, 05508-090, S\~ao Paulo, SP, Brazil
\and   Departamento de F\'isica Te\'orica y del Cosmos, Universidad de Granada, Facultad de Ciencias (Edificio Mecenas), E-18071 Granada, Spain
\and   Instituto Universitario Carlos I de F\'isica Te\'orica y Computacional, Universidad de Granada, E-18071 Granada, Spain\\}

   \date{Received June 4, 2015; accepted August 24, 2015}

\abstract{We use data from the Multi-Unit Spectroscopic Explorer (MUSE), recently commissioned at the {\it Very Large Telescope} (VLT), to study the kinematics and stellar population content of NGC 4371, an early-type massive barred galaxy in the core of the Virgo cluster. We integrate this study with a detailed structural analysis using imaging data from the Hubble and Spitzer space telescopes, which allows us to perform a thorough investigation of the physical properties of the galaxy. We show that the rotationally supported inner components in NGC 4371, an inner disc and a nuclear ring -- which, according to the predominant scenario, are built with stars formed from gas brought to the inner region by the bar -- are vastly dominated by stars older than 10 Gyr. Our results thus indicate that the formation of the bar occurred at a redshift of about $z=1.8^{+0.5}_{-0.4}$ (error bars are derived from 100 Monte Carlo realisations). NGC 4371 thus testifies to the robustness of bars. In addition, the mean stellar age of the fraction of the major disc of the galaxy covered by our MUSE data is above 7 Gyr, with a small contribution from younger stars. This suggests that the quenching of star formation in NGC 4371, likely due to environmental effects, was already effective at a redshift of about $z=0.8^{+0.2}_{-0.1}$. Our results point out that bar-driven secular evolution processes may have an extended impact in the evolution of galaxies, and thus on the properties of galaxies as observed today, not necessarily restricted to more recent cosmic epochs.}


\maketitle

\section{Introduction}
\label{sec:intro}

Evolved disc galaxies are characterised by an extended outer disc, dynamically supported by the rotation of its stars around the galaxy centre. These discs often present non-axisymmetric structures such as spiral arms or bars. The inner region of disc galaxies can contain a number of other components: e.g. a dynamically hotter spheroid, or disc-like structures, such as inner discs and nuclear rings\footnote{See \citet{But13} for an extensive review on galaxy morphology; see also \citet{Gad12} for a list of possible stellar structures in disc galaxies.}. Current scenarios attribute external and violent formation processes to the former -- such as mergers -- whereas inner disc-like structures are believed to have formed in internal and less disturbing processes. These processes are due to the impact of the non-axisymmetric components -- particularly bars -- on the otherwise axisymmetric potential of the disc. Those merger-built spheroids are sometimes named classical bulges, while the internal processes driven by non-axisymmetric components are sometimes labelled as part of secular evolution \citep[see e.g.][]{BinMer98,Elm98,Ath13,Kor13,Sel14}. Secular evolution is expected to be the dominant set of processes in galaxy evolution after the peak of star formation activity that occurs between redshifts 1 and 2, i.e. between about 8 and 10 Gyr ago. Before and during this intense period in cosmic history, interactions between galaxies are believed to be the dominant process through which galaxies evolve. After it, in generic terms, for galaxies outside dense environments such as clusters or groups, begins a less agitated period when internal secular evolution processes become important, particularly those driven by bars formed in galaxy discs.

One of the most important consequences of secular evolution is the inflow of gas through the bar to the central regions of galaxies, which results in the building of inner stellar components, a process seen in both theoretical and observational work \citep[see e.g.][and references therein]{ComGer85,Ath92b,SelWil93,PinStoTeu95,SakOkuIsh99,RauSal00,RegTeu03,RegTeu04,Ath05b,SheVogReg05,Woz07,Kna07,Gad09a,KimSeoSto12,ColDebErw14,EmsRenBou15,SorBinMag15}. The inflow of gas may originate inner discs and/or nuclear rings -- comprised by recently formed stars -- at or near the inner Lindblad resonance (ILR) associated to the bar, or rather at the region dominated by the bar x$_2$ orbits. A number of studies have found nuclear rings associated to bars \citep[from the pioneering works of \citealt{SerPas65,SerPas67} to e.g.][see also \citealt{ButCom96} and \citealt{MaoBarHo01}]{vanEms98,FalBacBur06,SarFalDav06,KunEmsBac10,PelKutvan12}, many of those rings showing intense star formation activity when cold gas is available. Other rings just evolve passively if most of the funnelled gas has been transformed into stars and there is no additional fuel for more star formation. Recent studies show that both the current and past star formation activity at the centres of barred galaxies are enhanced when compared to unbarred galaxies \citep[see e.g.][]{EllNaiPat2011,CoeGad11}, although not all barred galaxies conform to this picture. [In addition,  different studies have uncovered diverse results, likely because of differences in sample selection \citep{CacSanGor14,ZhoCaoWu15}. For example, \citet{CheConAth15} found ``no significant differences in the stellar populations of the bulges or the gradients of barred versus unbarred quiescent disk galaxies''].

Because these inner components are built from material brought from the disc (which is already dynamically and structurally mature enough to develop a bar), they are expected to show exponential, or near exponential, light profiles, as discs. In order to differentiate these inner structures from the classical concept of bulges residing in the centres of disc galaxies, a new nomenclature has developed, leading to terms such as pseudobulges, disc-like bulges or discy pseudobulges \citep[][see also \citealt{KorBar10}, \citealt{MenDebCor14} and \citealt{ErwSagFab14} for studies on the coexistence of different inner structures in a single galaxy]{KorKen04,Ath05b,DroFis07,FisDro10,Gad12,CheAthMas13}. Much work on these inner, secularly-built components and their relation to classical, merger-built bulges is still ongoing, particularly on their place in a hierarchical galaxy formation scenario \citep[see][and references therein]{Kor15,San15,Bou15,Fal15}.

Placing secular evolution in the hierarchical paradigm of galaxy formation is thus an important topic of research in extragalactic astrophysics today. Since bars are the major drivers of internal secular evolution \citep[and are common in the local universe, see e.g.][]{EskFroPog00,MenSheSch07,MasNicHoy11}, a major step forward would be to set the epoch in the evolution of the universe in which bars first ignited secular evolution processes. Ideally, one would like to be able to state for how long is the bar in a given barred galaxy influencing its host's evolution. The age of the stellar population in a bar is not necessarily a measure of the age of the bar. A young bar, i.e. one that has been recently formed, could very well be composed of old stars. Thus, different approaches have to be developed to address this problem. Progress has been slow from an observational viewpoint \citep[given the tremendous difficulties faced; see e.g.][]{GaddeS05,GaddeS06,PerSanZur07,PerSanZur09,SanOcvGib11,SeiCacRui15}. However, modern simulations at least indicate that bars, once formed, are difficult to dissolve (in the absence of major mergers), unless the disc is extremely gas-rich \citep{AthLamDeh05,BouComSem05,BerShlMar07,KraBouMar12}.

A powerful way to investigate the onset of bar-driven secular evolution -- from a statistical viewpoint -- is to study how the fraction of barred galaxies evolves with redshift. After the pioneering work by \citet{AbrMerEll99}, most of the latest studies have used images from the Advanced Camera for Surveys (ACS), onboard the {\it Hubble Space Telescope} (HST), taken with filters such as the F814W and F850LP, which correspond to broadband $I$ and SDSS $z$, respectively, i.e. as red as possible but still within the optical spectrum \citep[see][]{JogBarRix04,SheElmElm08,camcaroes10,MelMasLin14}. While the results are not entirely compatible between the different studies, there seems to be a consensus in that the fraction of disc galaxies with bars declines rapidly and monotonically from the local universe to $z\sim0.8$. \citet{SheMelElm12} suggest that this is due to an evolution in the dynamical state of discs. With time, discs grow massive enough and kinematically cold enough to develop bars. As alerted by \citet{SheElmElm08}, one major difficulty in studying the fraction of barred galaxies at $z\gtrsim0.8$ with the red ACS filters is that they correspond to rest-frame wavelengths in the ultra-violet part of the spectrum, where bars become exceedingly difficult to detect \citep[see][]{gilboimad07}. To avoid this issue, \citet{SimMelLin14} have used images taken with the infrared channel of the HST WFC3 (Wide Field Camera 3), with filters such as the F160W, allowing them to investigate the fraction of barred galaxies at $z>1$. Consistently with previous results, they found that the fraction of barred galaxies drops to $z\sim1$, but then stays approximately constant at about 10 per cent up to $z\sim2$. The images used by Simmons et al. are from the Cosmic Assembly Near-Infrared Deep Extragalactic Legacy Survey \citep[CANDELS;][]{GroKocFab11,KoeFabFer11}. These works then indicate that, statistically, bar-driven secular evolution became important at $z<1$. But the work of Simmons et al. also suggests that for a number of galaxies these processes have started earlier.

To understand the formation and evolution of galaxies, it is also necessary to understand when galaxies form stars and when galaxies cease star formation activities. While there are difficulties in understanding how galaxies in the field stop forming stars \citep[see e.g.][]{MarBouTey09}, in dense galaxy environments, such as galaxy clusters, the removal of cold gas from galaxies and their surroundings is expected to be efficient in quenching star formation \citep[see][]{BalBalNic04}. The formation of galaxy clusters is a long process that has initial phases at redshifts larger than about 3, and, although massive clusters have been observed at redshifts about unity or more \citep{BroRueAde10,BroStaGon13,ProBrePhi15,MaSmaSwi15}, it is expected that cluster formation is a process that goes on well below $z\approx1$ \citep{KraBor12,PlaSchByk14}. In fact, the nearby Virgo cluster is known for not being dynamically fully relaxed yet \citep{Bin99}. One thus expects that the effects of the cluster environment on the evolution of its constituent galaxies may evolve with time.

In this paper, we explore the capabilities of the recently commissioned MUSE (the Multi-Unit Spectroscopic Explorer, \citealt{BacAccAdj10}) integral field spectrograph -- installed on the European Southern Observatory {\it Very Large Telescope} (VLT) atop Cerro Paranal, Chile. MUSE is unique in its combination of field size, sensitivity, spatial resolution, wavelength range and spectral resolution. We use MUSE to study the inner $1'$ squared of the nearby galaxy NGC 4371. We use imaging data from the Spitzer and Hubble space telescopes, and, with MUSE, we are able to study in fine detail the inner components of the galaxy, and put constraints to their formation histories.

%In particular, we find that the nuclear ring --  at a radius of approximately $10''$ from the galaxy centre -- is dominated by a stellar population with mean age above about 10 Gyr. This puts a strong constraint to the cosmic epoch in which the bar formed, since the stars in the ring were born after the bar formed, which thus happened most likely at redshifts larger than unity. The old stellar ages we find throughout the inner $1'$ squared of NGC 4371 also suggest that the removal of available gas for star formation by the cluster environment in Virgo was already efficient when the cluster was at an early formation stage.

In the next section, we briefly describe the MUSE observations and the corresponding data reduction. In Sect. \ref{sec:4371}, we provide a general descripition of NGC 4371 and its environment. We perform in Sect. \ref{sec:struct} a detailed study of its structural properties, as derived using Spitzer images at 3.6$\mu$m -- from the Spitzer Survey of Stellar Structures in Galaxies \citep[S$^4$G,][]{shereghin10} -- and HST images, with their exquisite image quality. This structural analysis will prove fundamental in our understanding of the information extracted using MUSE. The extraction and analysis of kinematical information from the MUSE data cube is described in Sect. \ref{sec:kin}, while in Sect. \ref{sec:pops} we describe our study on the ages and metallicities of stars in the MUSE field. In Sect. \ref{sec:discuss} we discuss our results in the context of bar-driven secular evolution and environmental effects in galaxy clusters. Finally, Sect. \ref{sec:conc} summarises our main conclusions. Throughout this study we use a Hubble constant of ${\rm H_0}=67.8\,\rm{km}\,\rm{s}^{-1}\,\rm{kpc}^{-1}$ and $\Omega_{\rm m}=0.308$ in a universe with flat topology \citep{AdeAghArn15}.

\section{Observations and data reduction}
\label{sec:data}

NGC~4371 was observed as a science verification (SV) program [60.A-9313(A), PI: D.~A.~Gadotti] for MUSE during the nights 25th and 29th of June, 2014. MUSE covers an almost square $\sim 1'\times 1'$ field of view 
with a contiguous sampling of $0.2''\times0.2''$, corresponding to a massive dataset of about 90\,000 spectra
per pointing. A spectral coverage from 4750\AA\, to 9300\AA\, is achieved in the normal instrument setup at a spectral
resolution of $\approx2.5$\AA. The observations were performed under a seeing with full width at half maximum (FWHM) varying from $0.8''$ to $0.9''$. Sky transparency was with thin cirrus during the first night and photometric during the second night. Since we do not aim at using information sensitive to the flux calibration of the data, any uncertainty resulting from the thin cirrus in the first night is irrelevant here. Nevertheless, the presence of thin cirrus may limit the accuracy of the background subtraction.

We targetted the central $\sim 1'\times 1'$ region of NGC~4371 with MUSE (see Fig. \ref{fig:s4gimg}). The observations were distributed in two 1\,h observing blocks
with a total integration time of 1\,h on source. As the galaxy fills the MUSE field, we split the exposures per observing block into 3 exposures of 600\,s, and monitored the sky background by observing a blank sky field 
for 180\,sec following a SKY-OBJ-SKY-OBJ-OBJ-SKY pattern. To be able to reduce the effects of bad pixels and flat-fielding uncertainties we 
applied a small dither ($<1''$) of the field centre and rotated the entire MUSE field by 90${}^\circ$ between each object observation. Frames were taken to correct the exposures for bias and dark current, to flat-field the exposures, and to perform illumination correction. Wavelength calibration is achieved through a set of different arc lamp frames, and the exposures were flux-calibrated through the observation of a spectrophotometric standard star. Finally, the exposures were also finely registered astrometrically. All these frames were observed as part of the MUSE standard calibration plan, and we applied the calibration frames taken closest in time to the execution of our observing blocks.

The MUSE pipeline (version 0.18.2) was used to reduce the dataset. A detailed description of the pipeline
will be presented in Weilbacher et al. (in prep.), but we briefly outline the reduction process 
here. The process is split up into two steps. In the first one, each science frame is calibrated separately to take out instrumental effects. This
consists in subtracting the bias and dark current levels, flat-fielding the images (including illumination correction), extracting the spectra of all slices, and
 performing the wavelength calibration. Those calibrated science frames are combined into the final data cube 
in the second step. It includes the process of flux calibration, sky subtraction adopting a model sky spectrum computed
from the sky field, astrometric registration, correction for differential atmospheric refraction, and 
resampling of the data cube based on the drizzle algorithm \citep[see][]{WeiStrUrr12}.

\section{Introducing NGC 4371: general properties and environment}
\label{sec:4371} 

   \begin{figure}
   \centering
   \includegraphics[width=\hsize]{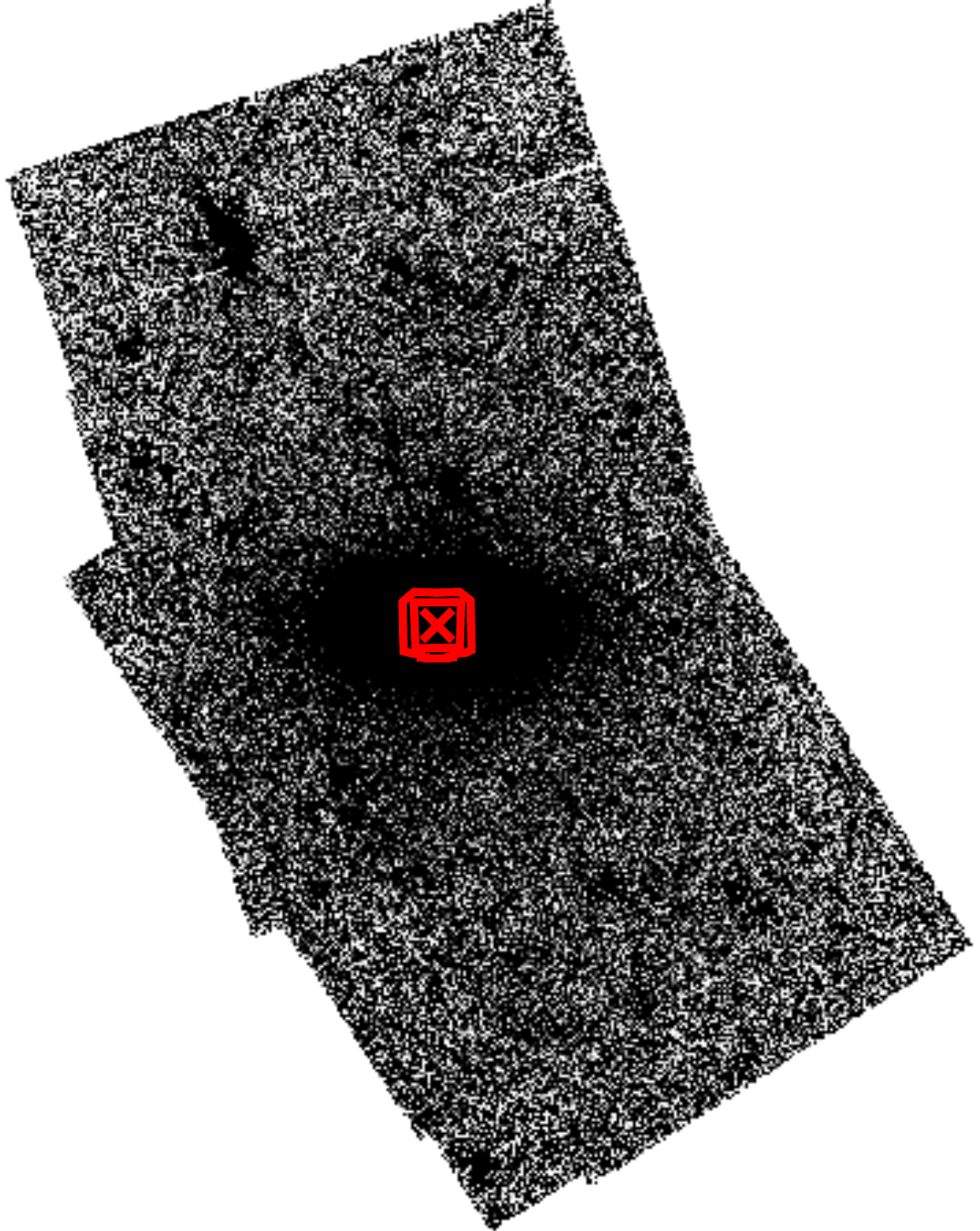}
   \vskip -1.5cm
   \includegraphics[width=0.7\hsize]{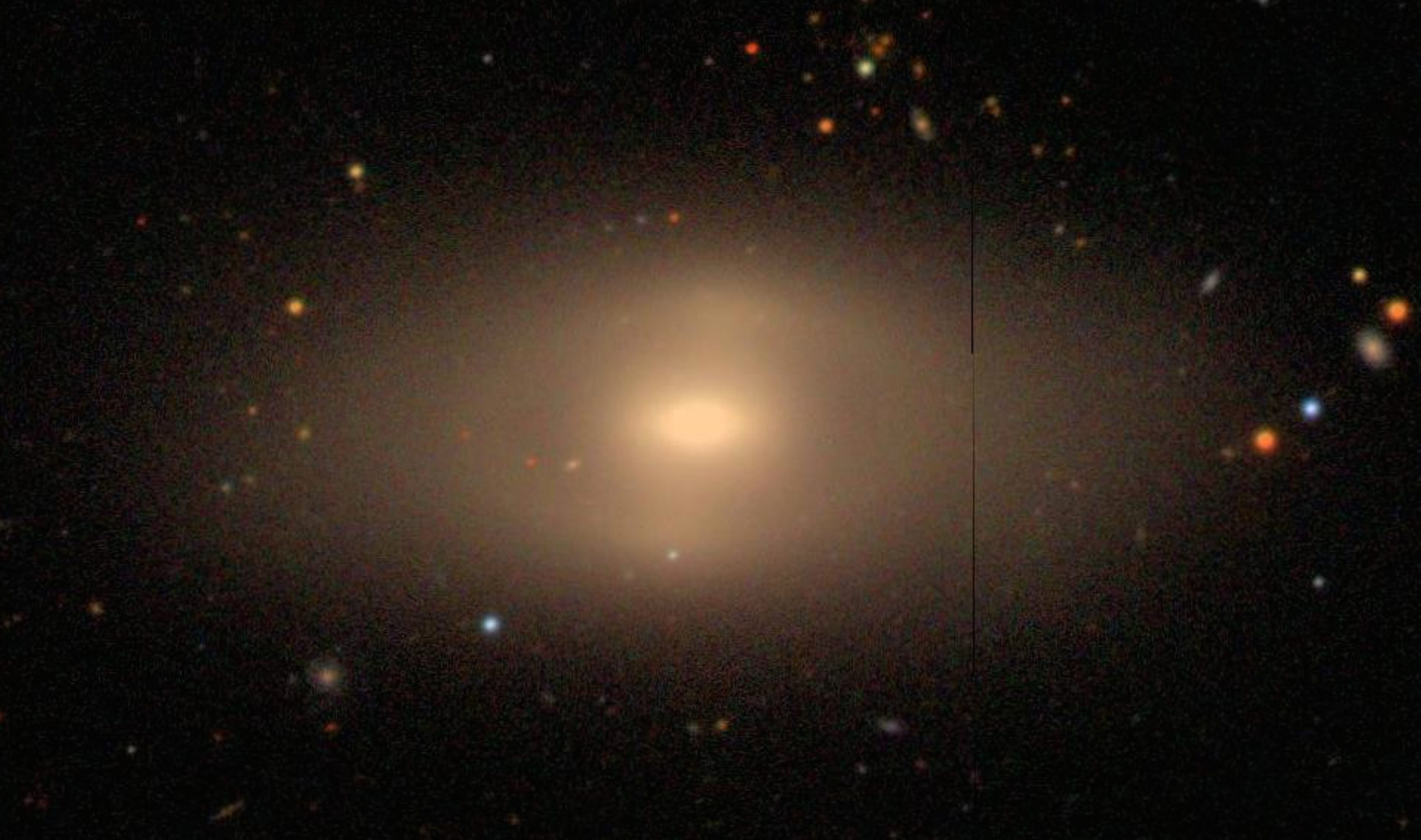}
   \includegraphics[width=0.49\hsize]{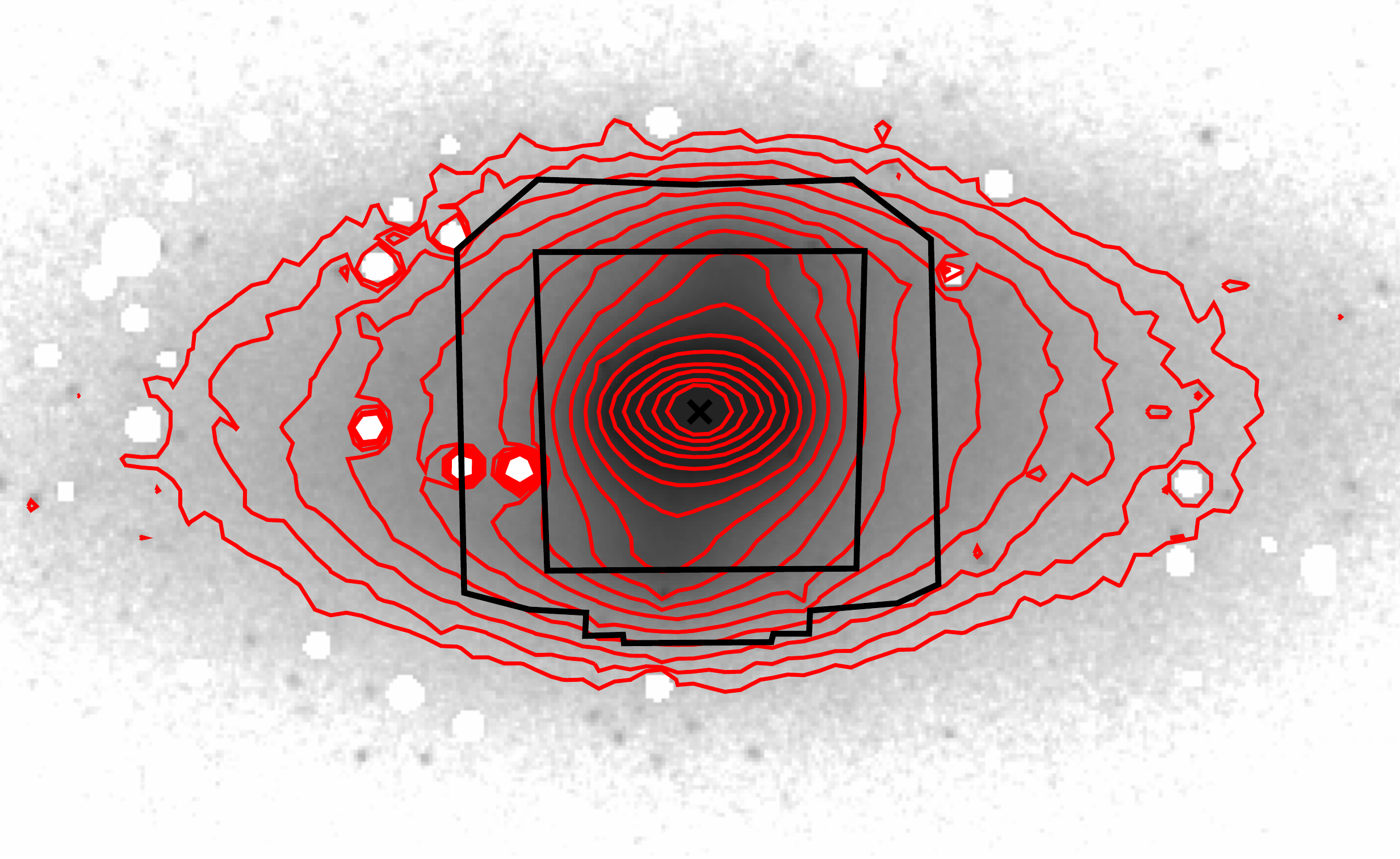}
   \includegraphics[width=0.49\hsize]{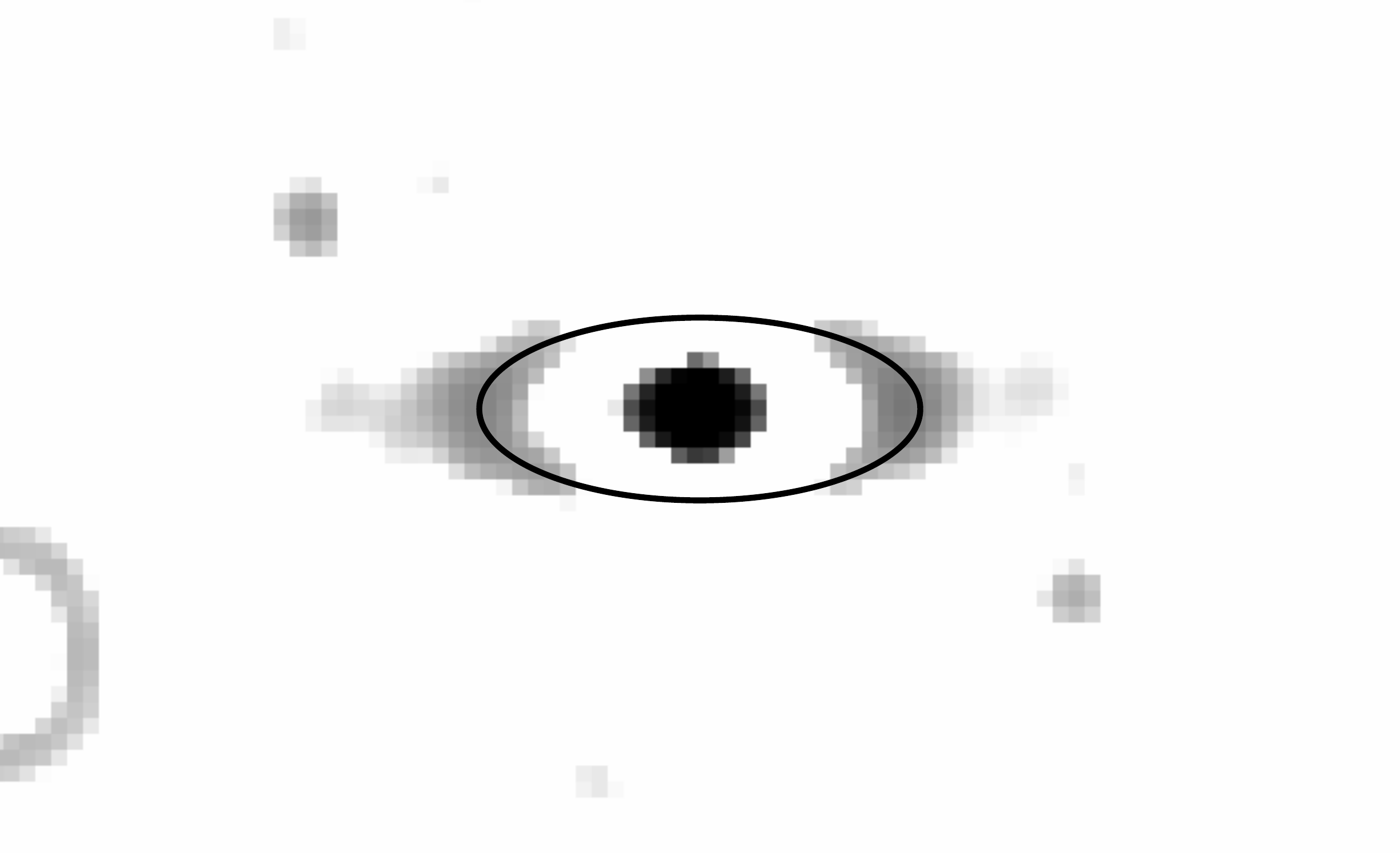}
      \caption{Top: reduced S$^4$G mosaic of NGC 4371 displayed using histogram equalization in order to highlight faint features. The inner red box represents the MUSE field of our exposures, with $1'$ on a side. This image suggests the presence of a very faint polar ring that could have originated from a satellite galaxy, but the evidence for it is weak at best. No sign of a violent recent interaction is seen. Middle: SDSS colour composite of NGC 4371, $6.6'$ on its longer side. Bottom left: central region of the S$^4$G image with isophotal contours overlaid in red and a diagram overlaid in black representing our MUSE exposures. The MUSE field is the inner trapezoid; the outer polygon area is used to acquire point sources for the slow guiding system. The bar is clearly seen. Bottom right: image obtained via unsharp masking of the S$^4$G image, showing clearly the ring near the centre of NGC 4371. This ring has a semi-major axis of $\approx10.4''$ and a semi-minor axis of $\approx4.3''$. It has a width of about $2''$, and is delineated by the black ellipse. North is up, East to the left in all panels.}
         \label{fig:s4gimg}
   \end{figure}

   \begin{figure}
   \centering
   \includegraphics[width=0.9\hsize]{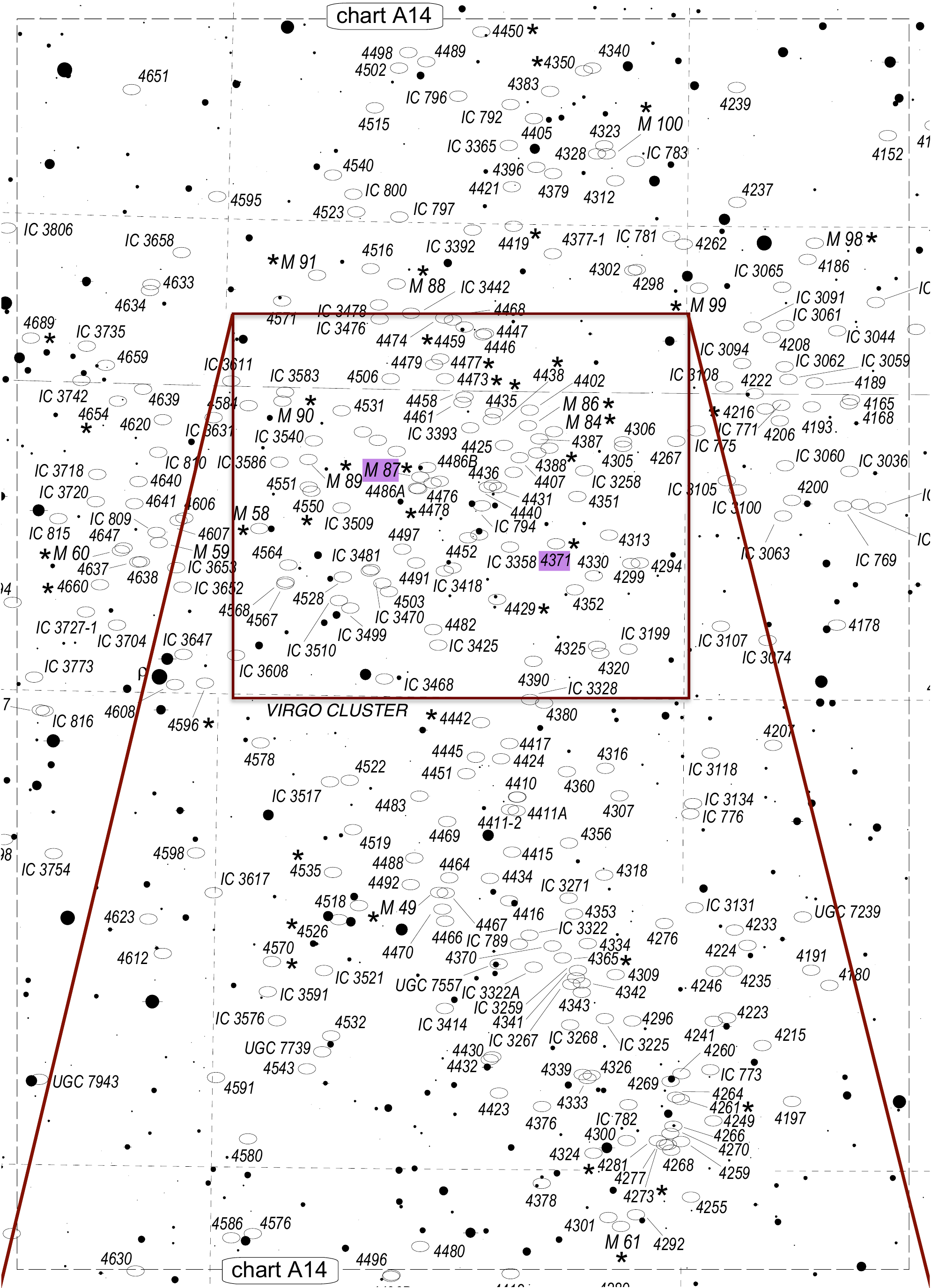}
   \vskip -0.1cm
   \includegraphics[width=0.9\hsize]{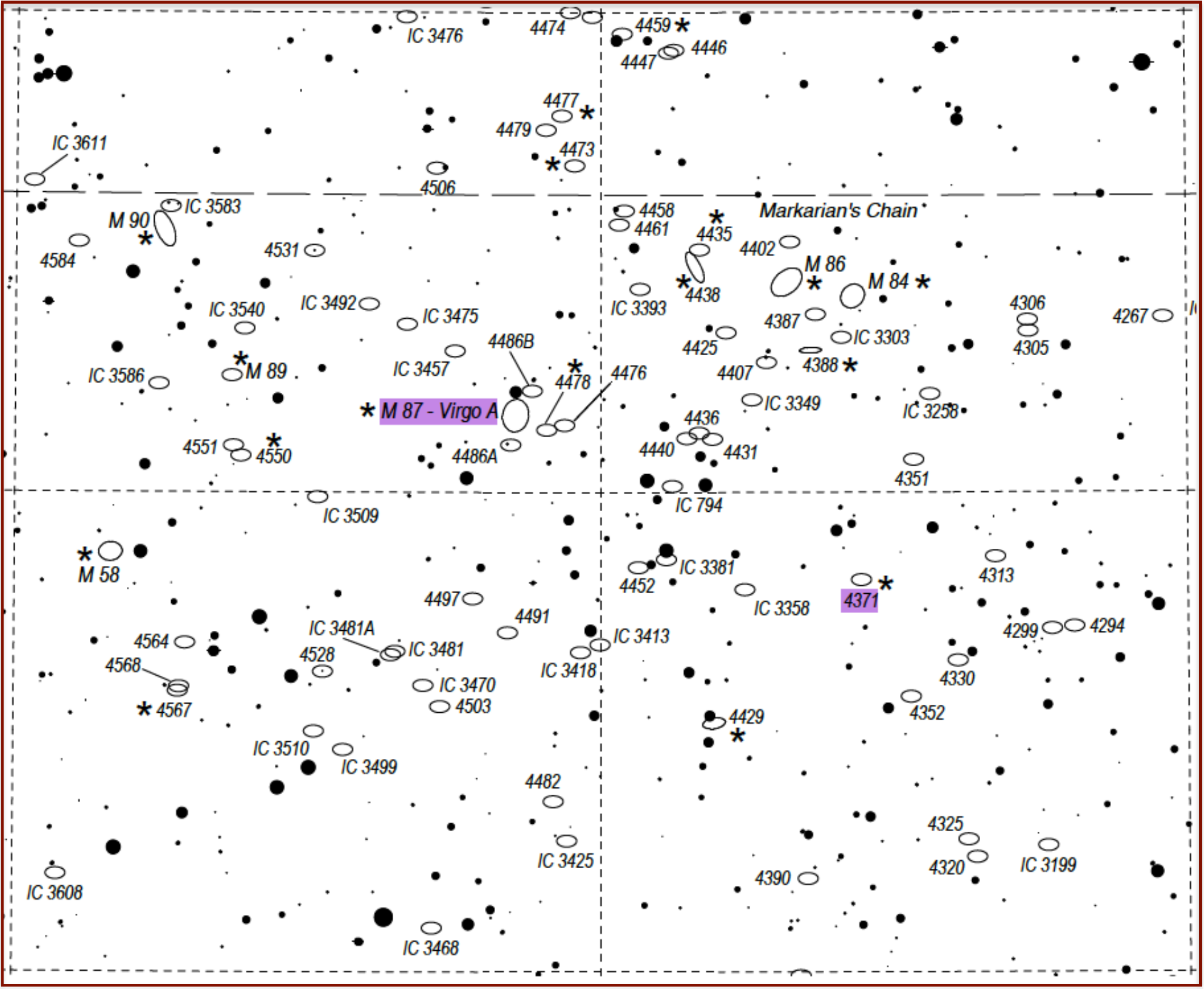}
      \caption{Top: sky map covering the Virgo cluster. The positions of NGC 4371 and M87 are highlighted, as well as an area of 16 squared degrees around the centre of the cluster. Bottom: expanded view of the 16 squared degrees box highlighted on the top panel. At the distance of the cluster, NGC 4371 is about 0.4 Mpc away from M87, which sits near the bottom of the cluster's potential well. {\it (Adapted with permission from the Deep Sky Hunter Star Atlas; see http://www.deepskywatch.com/ -- courtesy of Michael Vlasov).}}
         \label{fig:virgo}
   \end{figure}

NGC 4371 is an early-type galaxy that is in the core of the Virgo cluster. It is thus nearby ($d\approx16.9$ Mpc, \citealt{BlaJorMei09}), bright ($M_B\approx-19.2$, see HyperLeda\footnote{http://leda.univ-lyon1.fr/}; $M_{\rm AB, F850LP}\approx-21.2$,  \citealt{FerCotJor06}), massive ($M_\star\approx10^{10.8}\,{\rm M}_\odot$, \citealt{GalTreMar10}), and classified as a barred S0 (lenticular) galaxy \citep{ButSheAth15}. Buta et al. also mention that NGC 4371 has a nuclear ring, a barlens, an inner ring (i.e. with a radius close to the bar semi-major axis), as well as ansae at the bar ends, and an outer lens (see Fig. \ref{fig:s4gimg}). Ansae are stellar density enhancements found often in barred galaxies; they look like the bar handles \citep[see][for a recent study on ansae]{MarKnaBut07}. Barlenses have only recently been identified as an independent morphological feature. They are believed to be the face-on projection of box/peanuts \citep[][see also Fig. 8 in \citealt{GonGad15}]{LauSalAth14,AthLauSal14,Ath15}. In this case then, barlenses are just the inner parts of bars that are more extended than the outer parts in the vertical direction, as well as along the bar minor axis in the disc plane. As we will see below, our spectroscopic analyses do not reach far enough from the galaxy centre to include the outer lens, the ansae and the inner ring mentioned by \citet{ButSheAth15}. Therefore, these components will not be addressed here. Our MUSE field covers, however, part of the main disc, the bar, the nuclear ring and other inner components discussed in detail below, which hold important clues to the formation and evolution history of the galaxy.

The top panel of Fig. \ref{fig:s4gimg} shows the S$^4$G mosaic at an image stretch that highlights the faintest components of the galaxy. The sensitivity of this image is approximately ($1\sigma$) 27 AB mag arcsec$^{-2}$, or about 1 M$_\odot$ pc$^{-2}$. Yet there is no evidence for a violent recent interaction. On the other hand, it weakly suggests the presence of a very faint polar ring that could have originated from the encounter with a satellite galaxy. Therefore, it seems reasonable to assume that no dramatic merger has occurred recently in the galaxy. The SDSS colour composite in the middle panel shows clearly how uniform is the optical colour distribution across the galaxy, with no signs of recent star formation. The bottom left panel in Fig. \ref{fig:s4gimg} shows that our MUSE exposure covers almost the whole extent of the bar and a sizeable part of the major disc. The bottom right panel highlights a ring with a semi-major axis of $\approx10.4''$, semi-minor axis of $\approx4.3''$, and position angle similar to that of the major disc (see also Figs. \ref{fig:s4gprof} and \ref{fig:s4ggeo}). It has a width of about $2''$. This is the nuclear ring mentioned earlier, and we will call it the $10''$ ring.

In Fig. \ref{fig:virgo} we show the location of NGC 4371 in the Virgo cluster, as projected on the plane of the sky. NGC 4371 is a member of sub-cluster A \citep{GavBosSco99}. It is near the bottom of the cluster's potential well, and near one of its brightest galaxies, M87, and its extended X-ray halo of hot gas. The distance to NGC 4371, measured from surface brightness fluctuations, is $16.9\pm0.6$ Mpc \citep{BlaJorMei09}, while that to M87 is $16.4\pm0.5$ Mpc \citep[see][who arrive at this value by averaging different direct measurements]{BirHarBla10}. Projected on the plane of the sky, the distance between the two galaxies is about 1.5 degrees, corresponding to about 0.4 Mpc. The virial radius of the cluster is approximately 1.8 Mpc \citep{HofOlsSal80}, and the heliocentric radial velocity of NGC 4371, $933\,\rm{km}\,\rm{s}^{-1}$ \citep{CapEmsKra11}, is close to the mean Virgo velocity of $1035\,\rm{km}\,\rm{s}^{-1}$ \citep{MouHucFre00}. Altogether, these figures indicate that it is very probable that NGC 4371 undergoes significant interaction with the cluster environment. The results from e.g. \citet{PetWil04} and \citet{BroJanFlo11} indicate that the galaxy contains very little atomic and molecular gas. This could be the result of processes induced by the dense environment where the galaxy is located. We will come back to this point in Sect. \ref{sec:discuss}.

\section{Structural analysis}
\label{sec:struct}

In order to achieve as thorough as possible an understanding of the spatial variations in spectral properties in galaxies, such as kinematic measurements and stellar ages and metallicities, it is fundamental to study their structural properties. To put it differently, a very powerful way to investigate the formation histories of galaxies is to combine spectral information, such as kinematics, with photometric/morphological information, such as the structural properties of different stellar components in a galaxy. As an example, one would use photometric information to understand where the bulge dominates in a given galaxy and in this way be able to isolate -- using only spectra from that region -- the bulge mean stellar metallicity or age from the corresponding parameters of the disc or the galaxy as a whole \citep[see e.g.][]{CoeGad11}.
%Integral field spectroscopy is very advantageous in such kind of analysis, and MUSE, with its superb sensitivity and spatial resolution, is very powerful in this regard.
This section is thus mainly dedicated to investigate in detail the structural properties of NGC 4371 before we dive into the spectral properties derived with our MUSE SV data.

We take advantage of the fact that NGC 4371 was already observed with powerful imagers such as those onboard the {\it Spitzer Space Telescope} and the HST. We use the former to derive structural properties of the major disc and bar of the galaxy. We use a $3.6\mu$m image (from the Infrared Array Camera, IRAC, channel 1) obtained as part of the S$^4$G. The pixel size of the S$^4$G image is $0.75''$, and the PSF (point spread function) FWHM is $\approx1.8''$ \citep{KimGadShe14}. At these wavelengths, dust extinction and emission have little effect in our measures, so that they provide a very accurate representation of the bulk of the stellar population in the galaxy. \citet{MeiSchKna12} find that the contribution from PAHs (polycyclic aromatic hydrocarbons) and hot dust is between about 5 and 10 per cent of the integrated light at $3.6\mu$m. The HST image (downloaded from the Hubble Legacy Archive) is used to study the inner structural components, where the pixel size and spatial resolution of the Advanced Camera for Surveys (ACS) provide a precious sharp view. The pixel size is $0.05''$ and the PSF FWHM is $\approx0.09''$. Although in this case we have to use an optical passband, we use an image taken with the F850LP filter, which has the peak of transmission at about 850 nm, in order to minimise dust effects. This HST image was taken as part of the ACS Virgo Cluster Survey \citep{CotBlaFer06}.

\subsection{Major structures}

   \begin{figure}
   \centering
   \includegraphics[width=\hsize]{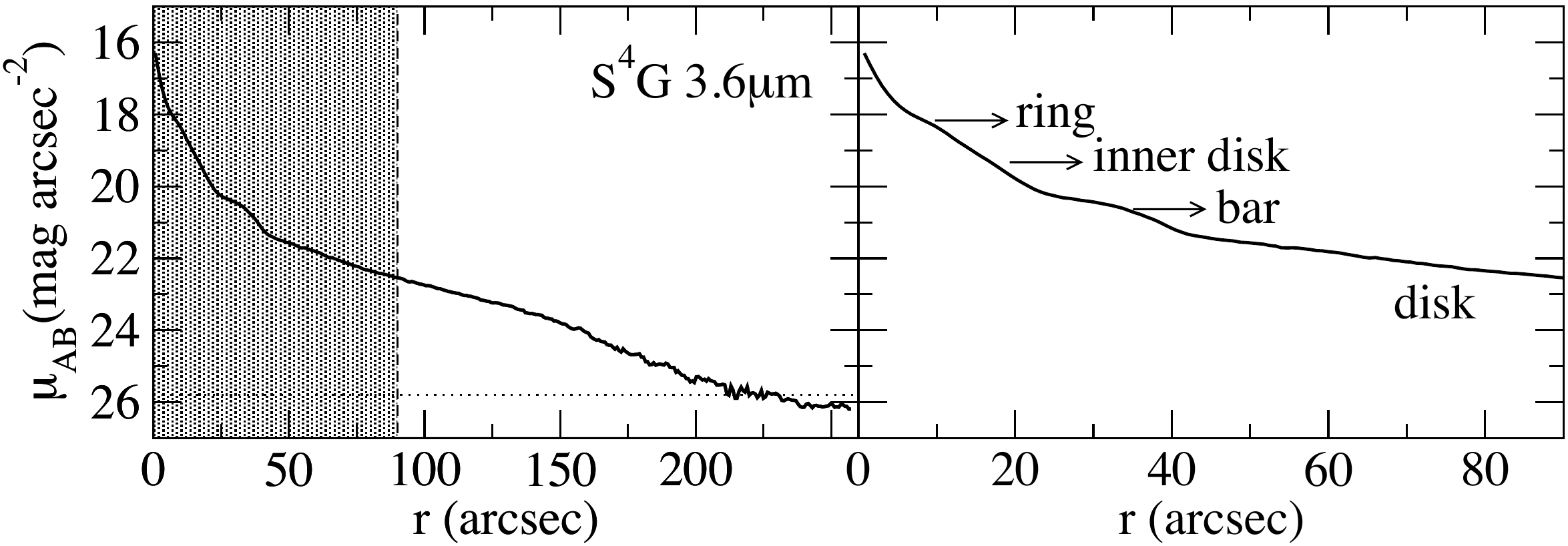}
   \includegraphics[width=\hsize]{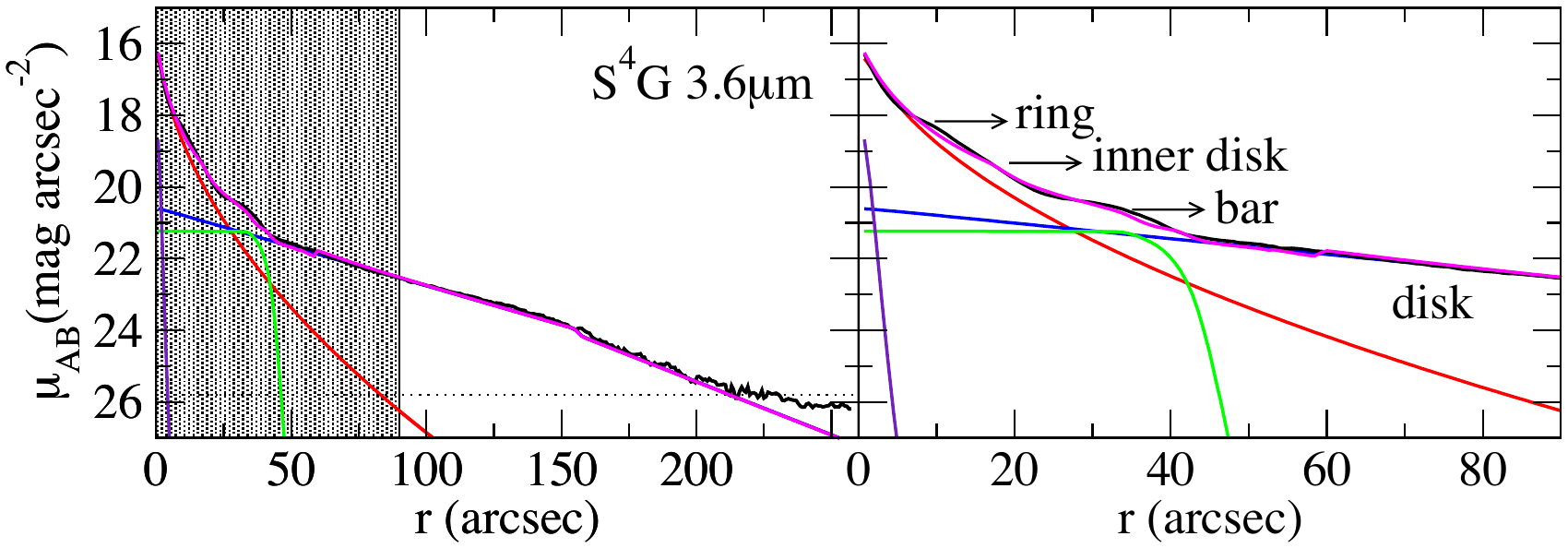}
      \caption{Surface brightness radial profile of NGC 4371. The top panels show only the profile derived from ellipse fits to the S$^4$G image. The bottom panels also show the profiles of the different model components obtained with {\tt BUDDA}: bulge in red, disc in blue, bar in green, and central point source in violet. The profile corresponding to the total {\tt BUDDA} model is in magenta. The horizontal dotted line marks the surface brightness level below which the background noise becomes important. The panels on the right focus on the inner 90" (shaded area on the left-hand panels). The outer exponential disc, as well as the bump caused by the bar are clearly seen. We also point out the bump at a radius of 10" produced by the ring shown in Fig. \ref{fig:s4gimg}, and a clearly exponential region that dominates the profile between the ring and the bar-dominated region, that appears to be an inner disc, given its exponential profile.}
         \label{fig:s4gprof}
   \end{figure}

   \begin{figure}
   \centering
   \includegraphics[width=0.646\hsize]{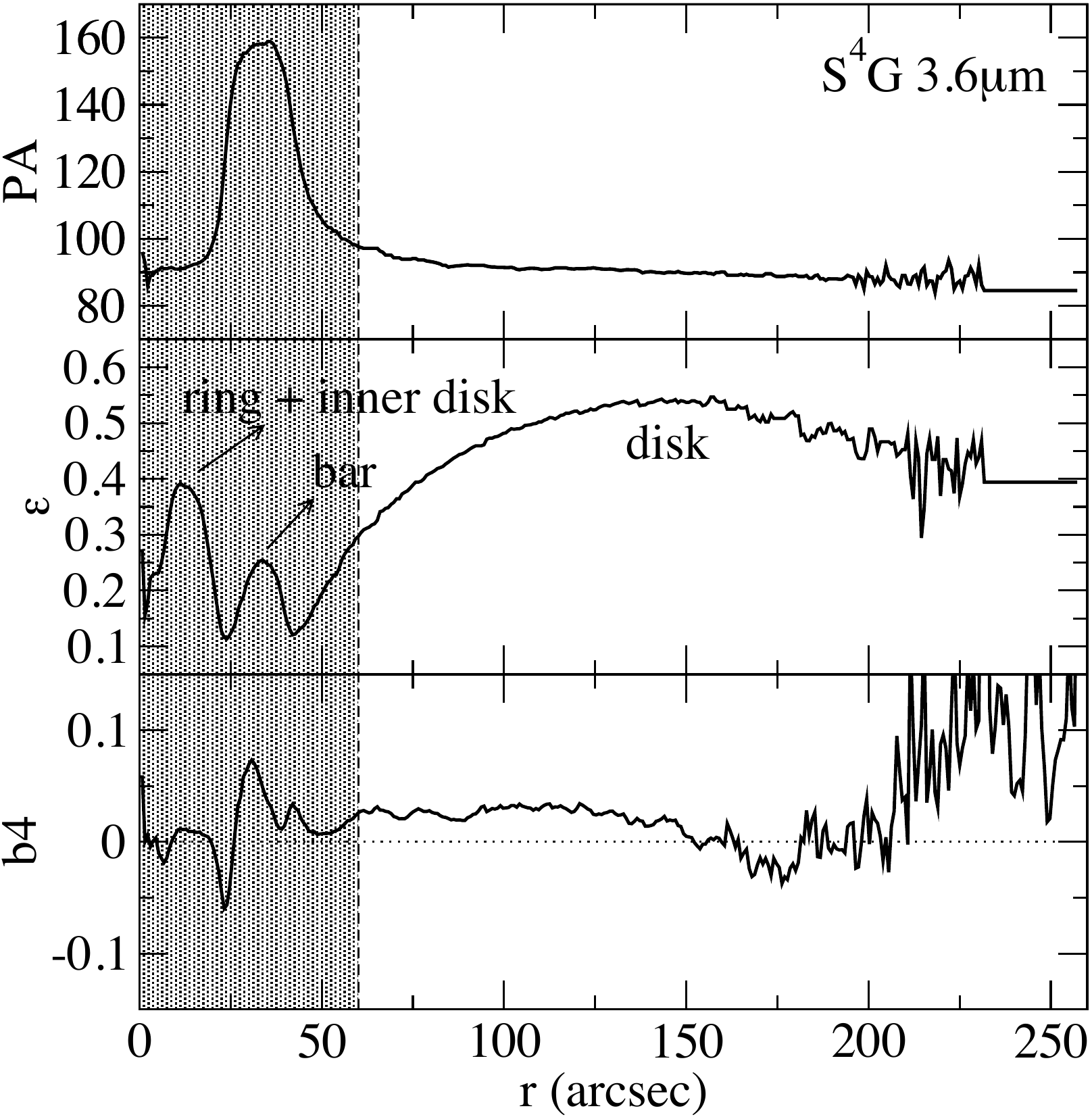}
   \includegraphics[width=0.345\hsize]{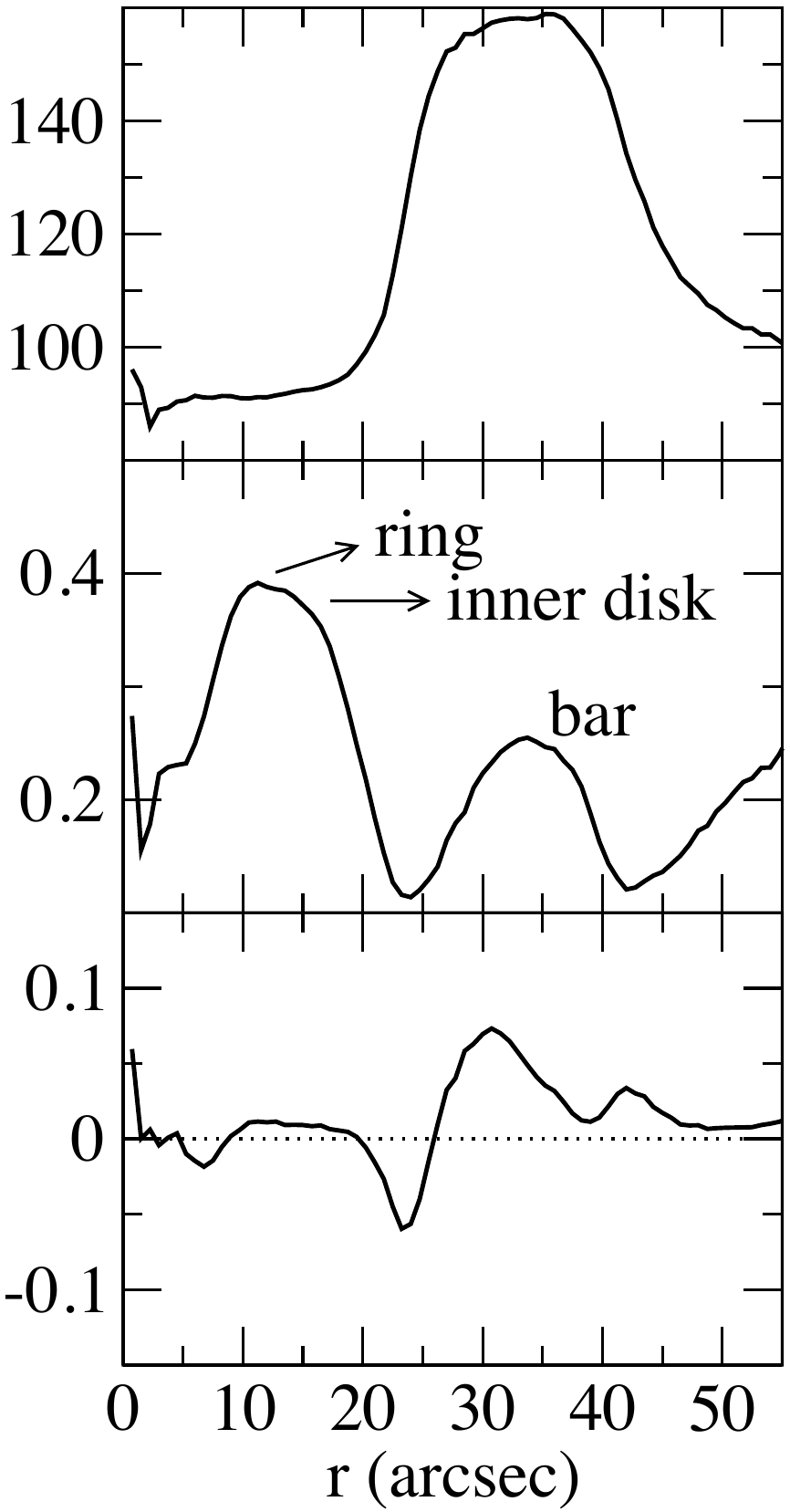}
      \caption{Radial profiles of position angle, ellipticity and the b$_4$ Fourier coefficient of the ellipses fitted to the isophotes of NGC 4371 in the S$^4$G image, as indicated. The right-hand panels focus on the inner $60''$ (shaded area on the left-hand panels). In the ellipticity profile we indicate features produced by the $10''$ ring, inner disc, bar and major disc.}
         \label{fig:s4ggeo}
   \end{figure}

\begin{table}
\caption{\label{tab:decomps1}Structural properties of the stellar components in NGC 4371 as derived from the S$^4$G image {\tt BUDDA} fit.}
\centering
\begin{tabular}{lcc}
%\hline\hline
%\\
Photometric Bulge\\
\hline 
effective surface brightness & $\mu_{e,{\rm b}}$                 & 19.1 \\
effective radius                     & $r_{e,{\rm b}}$                     & 11.6 \\
S\'ersic index                        & $n_{\rm b}$            & 2.0 \\
ellipticity                                & $\epsilon_{\rm b}$ & 0.26 \\
position angle                       & PA$_{\rm b}$         & 91 \\
\hline
\\
Disc\\
\hline
inner central surface brightness  & $\mu_{0,{\rm inn}}$  & 20.6 \\
inner scalelength                         & $h_{\rm inn}$           & 49.7 \\
ellipticity                                       & $\epsilon_{\rm d}$   & 0.48 \\
position angle                              & PA$_{\rm d}$           & 92 \\
break radius                                 & $r_{\rm br}$             & 156 \\
outer central surface brightness  & $\mu_{0,{\rm out}}$ & 19.5 \\
outer scalelength                         & $h_{\rm out}$           & 36.5 \\
\hline
\\
Bar\\
\hline
semi-major axis  & $L_{\rm bar}$            & 34.8 \\
S\'ersic index      & $n_{\rm bar}$            & 0.2 \\
ellipticity              & $\epsilon_{\rm bar}$ & 0.51 \\
position angle     & PA$_{\rm bar}$         & 159 \\
boxiness             & $c$                            & 2.7 \\
\hline
\\
Luminosity/Mass fractions\\
\hline
photometric bulge/total          & B/T                      & 0.427 \\
inner disc/total                       & D$_{\rm inn}$/T   & 0.437 \\
outer disc/total                       & D$_{\rm out}$/T  & 0.056 \\
bar/total                                 & Bar/T                   & 0.077 \\
central point source/total       & PS/T                    & 0.003 \\
\hline
\end{tabular}
\tablefoot{Luminosity parameters are in units of $3.6\mu$m AB mag arcsec$^{-2}$. Spatial measurements are in units of arcseconds. Position angles are in degrees from North Eastwards. Note that the fit also includes a central point source.}
\end{table}

The top panels of Fig. \ref{fig:s4gprof} show the radial surface brightness profile derived from the S$^4$G image with ellipse fits to the isophotes, using the {\tt ellipse} task in {\tt IRAF}. The bottom panels show the result of an image decomposition with {\tt BUDDA} \citep{deSGaddos04,Gad08}. The fit includes models for a photometric bulge, a bar and a type II disc \citep[see e.g.][and references therein, for the different types of disc profiles\footnote{We note that \citet{ErwPohBec08} do not find the type II disc break in NGC 4371 with a profile derived from ellipse fits with fixed position angle and ellipticity. We also do not find the break in a profile derived from a major-axis cut. That is the only difference between the latter and the results from our ellipse fits.}]{ErwPohBec08}, and it also takes into account the presence of a bright central point source. The bulge and the bar are fitted using S\'ersic functions, whereas the disc is fitted with a double exponential.
%The ring shown in Fig. \ref{fig:s4gimg} is also seen in a residual image (not shown here) obtained by subtracting from the S$^4$G image the {\tt BUDDA} model.
A signature of the ring shown in Fig. \ref{fig:s4gimg} can be seen in the surface brightness profile, i.e. the characteristic bump at a galactocentric radius of about $10''$.  One also sees a clearly exponential section in the surface brightness profile, between the ring and the bar, dominating to about $20''$, which thus appears to be an inner disc. We skip for now the discussion on whether this structure we are calling an inner disc is a barlens, but we note that barlenses usually also show exponential profiles. We will return to this question in Sect. \ref{sec:kin}, where the analysis of the kinematical properties of NGC 4371 will shed light on it. The model for the photometric bulge essentially accounts for all light above the inward extrapolation of the disc exponential profile, except the bar.

The profiles shown in Fig. \ref{fig:s4ggeo} were also derived from ellipse fits. They show the features in the isophotes' geometric properties that correspond to the major disc, the bar and the $10''$ ring. Since the ring has a width of only $\approx2''$, the plateau in ellipticity seen between $\approx11''$ to $\approx17''$ does seem to be associated with the inner disc. Assuming the major disc to be infinitely thin, and taking into account the global maximum in the ellipticity profile -- which happens to be at the region where the disc dominates -- one can derive the galaxy inclination with respect to the plane of the sky to be $\approx60^\circ$. The main structural parameters obtained from the fit to the S$^4$G image can be found in Table \ref{tab:decomps1}.

\subsection{Inner structures at high resolution}
   
The fit to the HST image was done with {\tt GALFIT} \citep{PenHoImp02,PenHoImp10}. This is because {\tt GALFIT} easily allows an arbitrary number of components to be included in the model, and this proved to be necessary when we studied the HST image, in order to account for the different structural components in the galaxy central region. It also allows a very careful modelling of the PSF, by using a PSF image derived directly from the ACS image, i.e. without the need to assume that the PSF follows any particular function. This will be important here, as below we will look at structural components at the very centre of NGC 4371. The structural parameters found for the major disc and the bar in the fit of the S$^4$G image were kept fixed in the fit of the HST image, of which we focus on the central components.

   \begin{figure}
   \centering
   \includegraphics[width=\hsize]{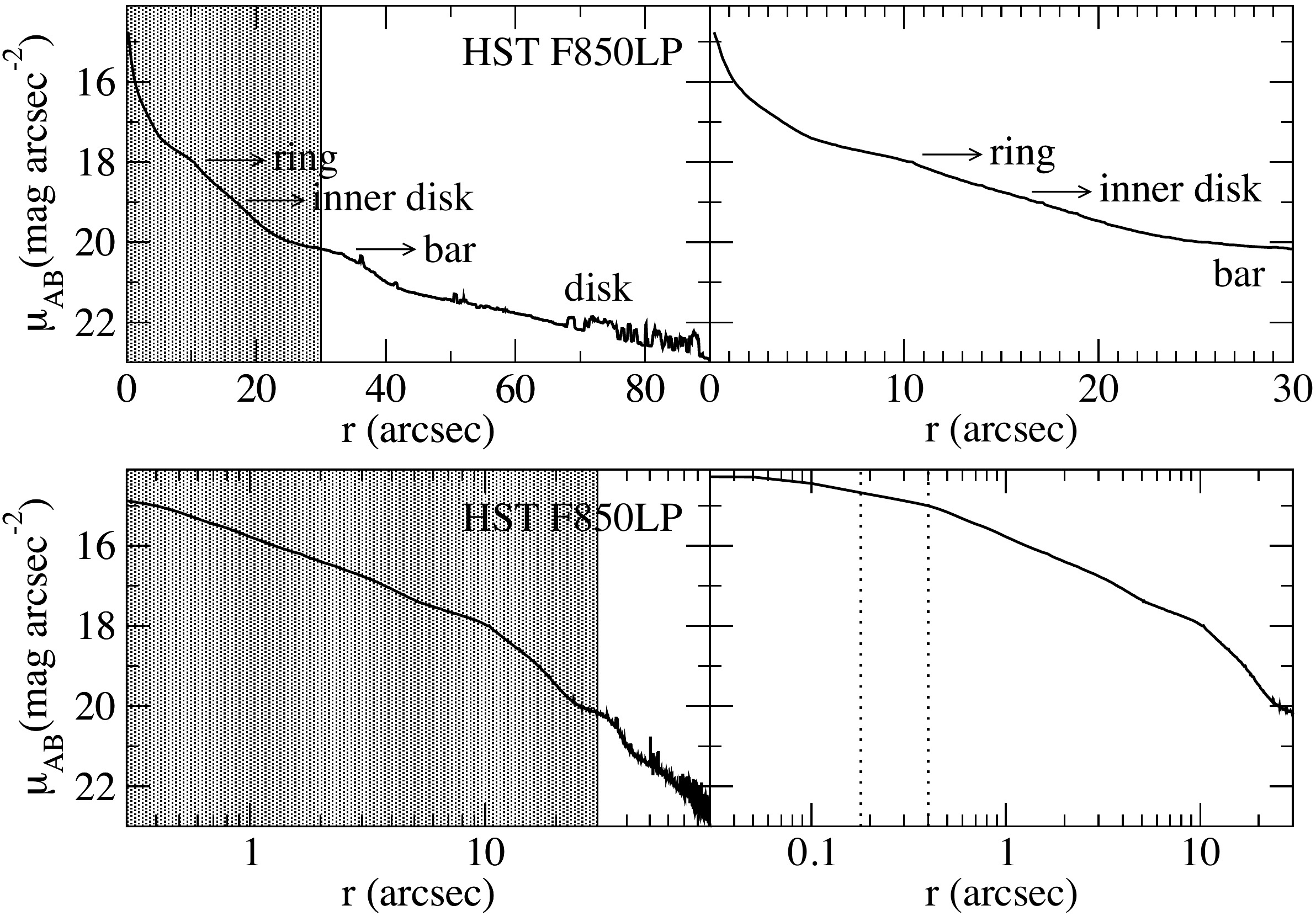}
      \caption{Surface brightness radial profiles of NGC 4371 derived from ellipse fits to the HST image. The right-hand panels focus on the inner $30''$ (shaded area on the left-hand panels). Bottom panels have the radial axis displayed on a log scale. The vertical dashed lines mark the positions of the nuclear features seen in the position angle and ellipticity radial profiles derived from the HST image (see Fig. \ref{fig:hstgeo}).}
         \label{fig:hstprof}
   \end{figure}

   \begin{figure}
   \centering
   \includegraphics[width=\hsize]{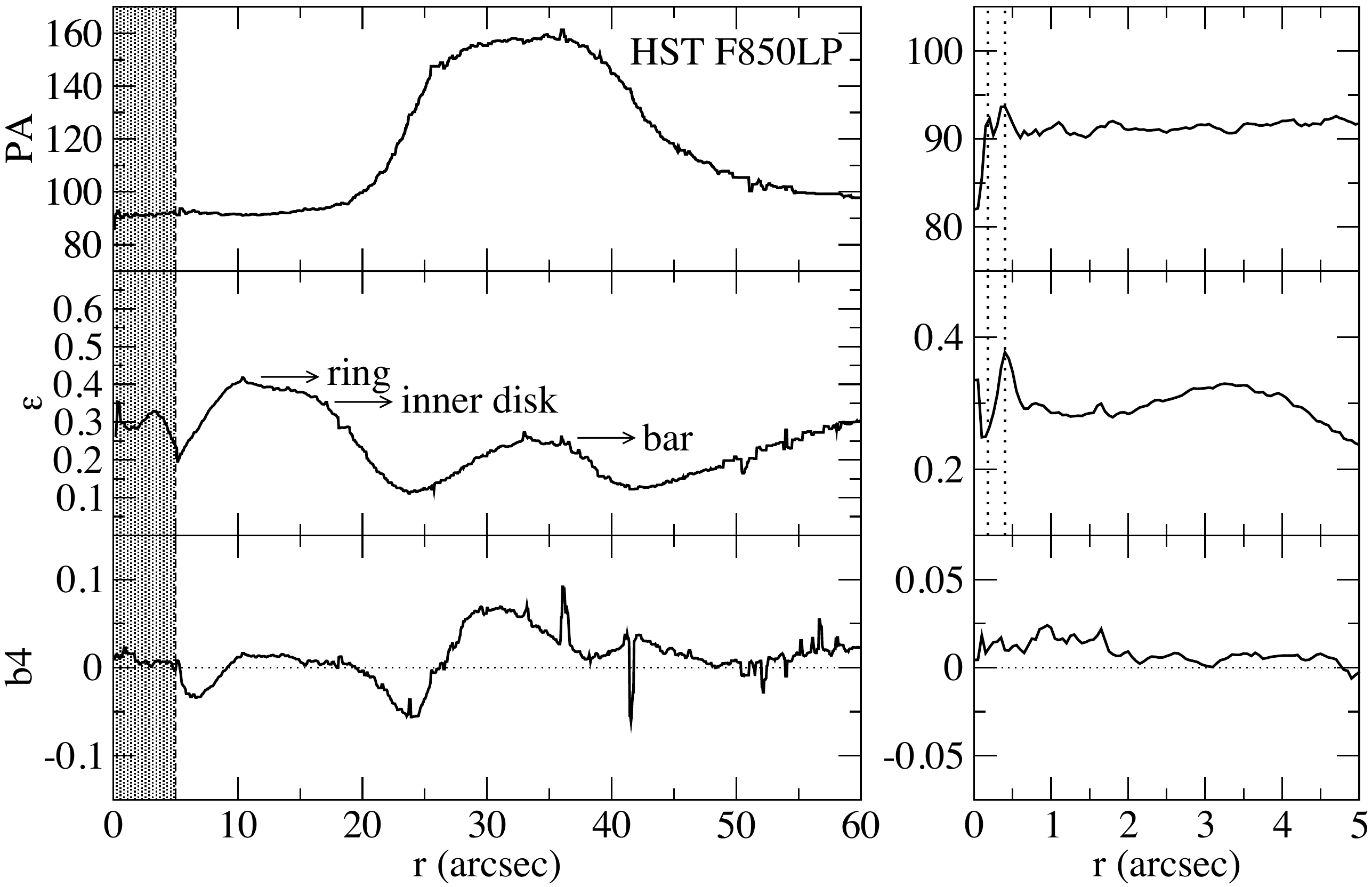}
      \caption{Radial profiles of position angle, ellipticity and the b$_4$ Fourier coefficient of the ellipses fitted to the isophotes of NGC 4371 in the HST image, as indicated. The right-hand panels focus on the inner $5''$ (shaded area on the left-hand panels). The positions of nuclear features seen in the position angle and ellipticity profiles are marked by vertical dashed lines.}
         \label{fig:hstgeo}
   \end{figure}

Figures \ref{fig:hstprof} and \ref{fig:hstgeo} show results from ellipse fits to the HST image. Given the improved spatial resolution with respect to the S$^4$G image, we can now focus on the galaxy central region. Features in the ellipticity and position angle radial profiles can be seen at $0.18''$ and $0.4''$ from the centre, in the right-hand panels of Fig. \ref{fig:hstgeo}. This figure also shows a peak in the ellipticity profile at about $3''$.

   \begin{figure}
   \centering
   \includegraphics[width=\hsize]{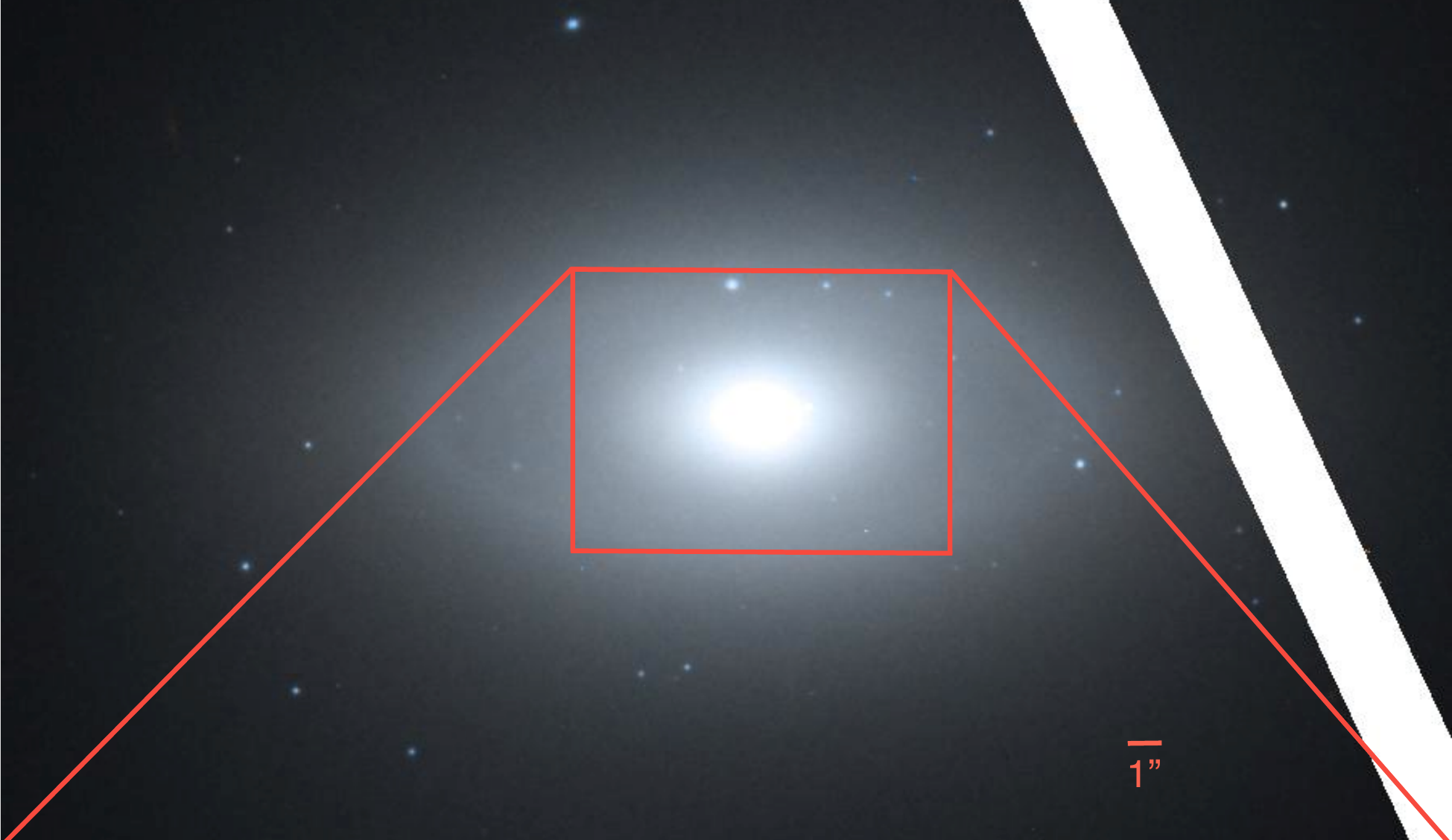}
   \vskip -0.3mm
   \includegraphics[width=\hsize]{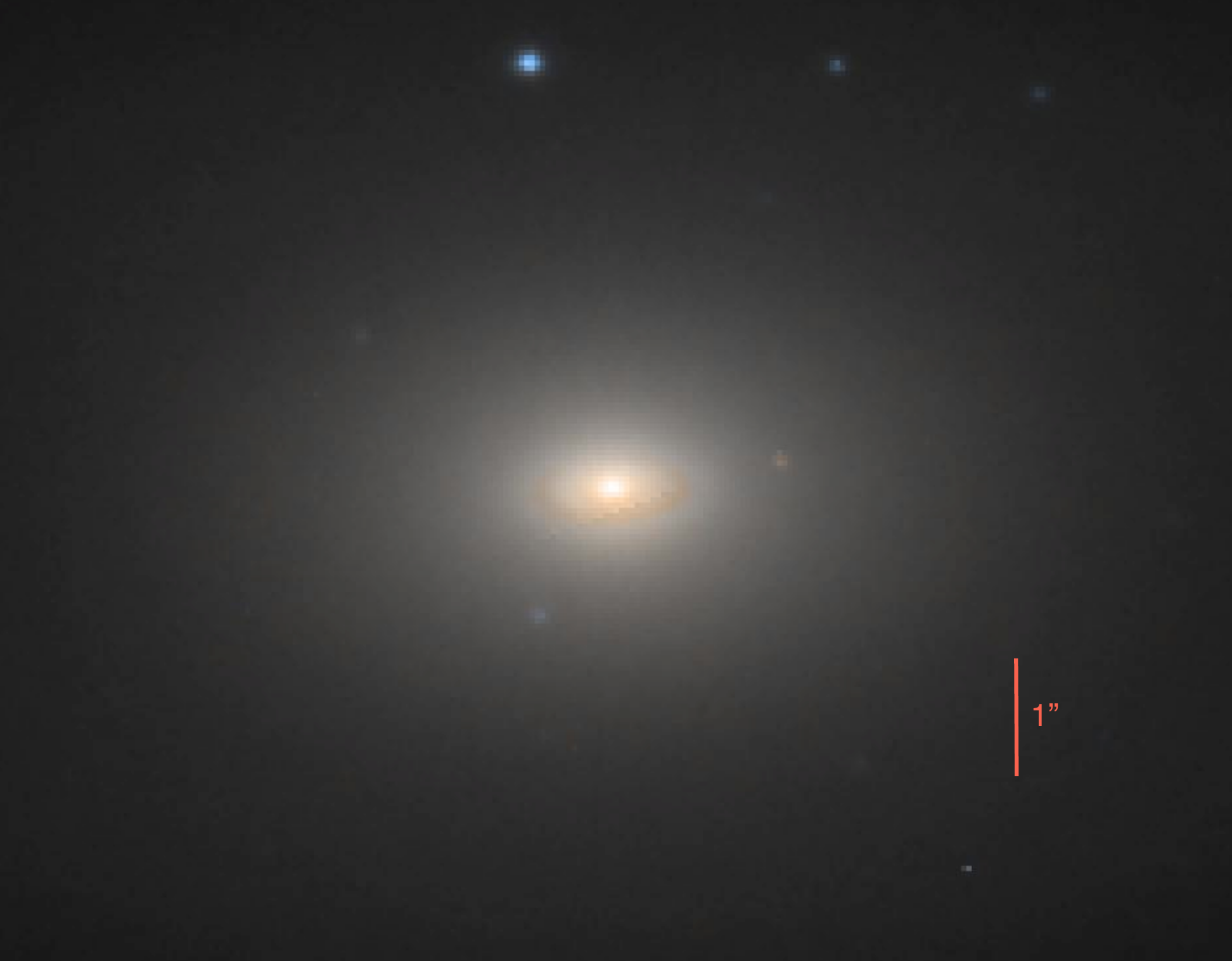}
      \caption{Colour composites produced from HST images (F850LP and F475W) at the Hubble Legacy Archive website. Top: a clear view of the $10''$ ring. Bottom: the very inner region of NGC 4371. The blue point source on the North is $\approx4''$ from the centre. The bottom panel clearly shows distinct features in the very centre, all elongated, with major axes positioned along the East-West direction. From inside out, a bright yellow-reddish, slightly less elongated structure with a semi-major axis of $\approx0.15''$. Further out, a dusty ring with semi-major axis of $\approx0.6''$, and just at the outer rim of it, at $\approx0.8''$, a bluish ring. These dusty disc-like structures were also noted by \citet[see also \citealt{ComKnaBec10}]{FerCotJor06}. North is up, East to the left.}
         \label{fig:hstcolor}
   \end{figure}

In Fig. \ref{fig:hstcolor} we take a close look at the stunning features shown with HST imaging near the centre of the galaxy. This figure shows a colour composite using F850LP and F475W HST images. The bottom panel zooms in the upper panel, focussing on the very central features. The $10''$ ring is again clearly seen in the upper panel. One can also see a few other features in the bottom panel. From inside out, starting from the very centre, a bright yellow-reddish, slightly less elongated structure with a semi-major axis of $\approx0.15''$. Further out, a dusty ring with semi-major axis of $\approx0.6''$. Both these features are at positions roughly coincident with the features seen in the position angle and ellipticity profiles (at $0.18''$ and $0.4''$, respectively, see Fig. \ref{fig:hstgeo}). Just at the outer rim of the dusty ring, at $\approx0.8''$, one can also see a bluish ring. These dusty disc-like structures were also noted by \citet{FerCotJor06} and \citet{ComKnaBec10}. We will call these structures (except for the less elongated, central component within a radius of $\approx0.15''$) collectively as the nucleus, and they are included as a single component in the HST image fit with {\tt GALFIT}.

   \begin{figure}
   \centering
   \includegraphics[width=\hsize]{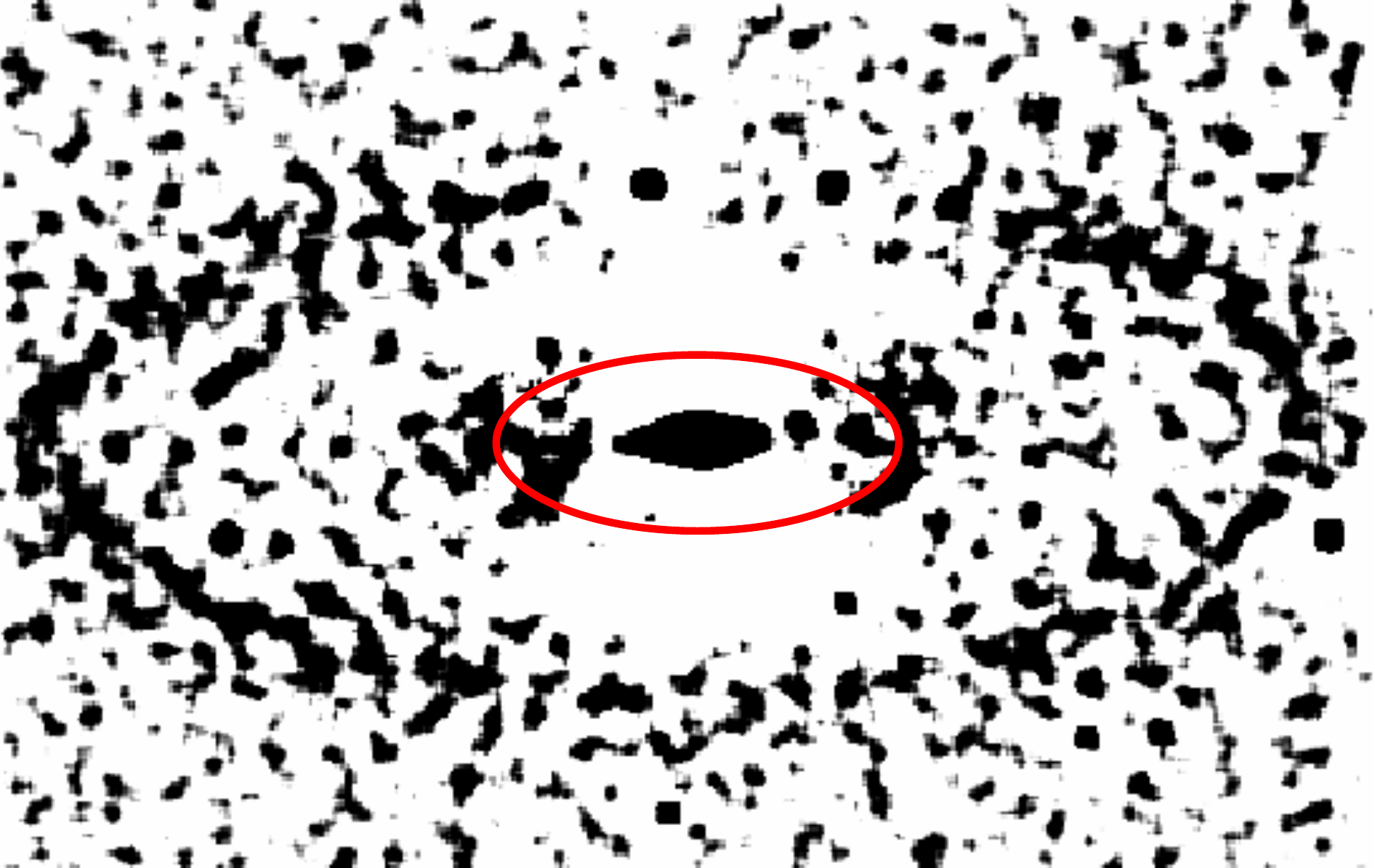}
      \caption{The central region of the F850LP HST image after unsharp masking. The $10''$ ring is clearly visible, as well as another elongated, disc-like structure with semi-major axis of $\approx3.2''$ and semi-minor axis of $\approx1.4''$, as delineated by the red ellipse. It has a width of about $2''$. The structures seen at the very centre of the galaxy in the bottom panel of Fig. \ref{fig:hstcolor}  (within $\approx0.8''$) are within the elongated dark blob at the centre of this figure.}
         \label{fig:hstum}
   \end{figure}

Circumscribing all these components there is a fainter elongated structure. It can also be seen in the unsharp mask shown in Fig. \ref{fig:hstum}. It has a semi-major axis of $\approx3.2''$ and a semi-minor axis of $\approx1.4''$. It has a width of about $2''$. The elongated dark blob seen at the centre in Fig. \ref{fig:hstum} corresponds to the structures seen in Fig. \ref{fig:hstcolor} within $\approx0.8''$.

The fit to the HST image thus includes models for the major disc, the bar, the inner disc, the nucleus, and again accounts for a central point source, i.e. the less elongated, central component within a radius of $0.15''$. Note that no extra component is necessary to account for the presence of a bulge. Also, the $10''$ ring and the $3''$ disc-like structure at the centre are not included in the fit. The discs are fitted with single exponential functions, whereas the bar is fitted with a Ferrer function, and the nucleus with a S\'ersic function. The fit extends to $118''$, before the disc break seen with the S$^4$G image, and thus a double exponential function for the major disc is not necessary. Results from this fit can be found in Table \ref{tab:decomps2}.
%Figure \ref{fig:resid} shows a residual image from the HST image fit. It corroborates the previous analysis and confirms the presence of the $3''$ component. It also shows that, even with a model component to fit the nucleus, a residual, very elongated structure remains at the central region.

In the top right panel of Fig. \ref{fig:hstprof}, one can see that the inner $\sim4''$ can be described as a pair of two different exponentials. The outer one starts at about $1''$ and extends to about $4''$, and thus seems to be associated to the $3''$ disc-like component. Within $1''$ there is another exponential section, therefore associated to the nucleus component. This is consistent with the S\'ersic index found for the nucleus, i.e. $n_{\rm n}=1.1$ (see Table \ref{tab:decomps2}), or essentially an exponential function. The central point source also makes its mark in the bottom right panel of Fig. \ref{fig:hstprof}. One sees that the curvature of the surface brightness radial profile changes right at the position of the inner vertical dashed line, which marks the nuclear feature seen in the position angle and ellipticity profiles (Fig. \ref{fig:hstgeo}).

\begin{table}
\caption{\label{tab:decomps2}Structural properties of the inner stellar components in NGC 4371 as derived from the HST image {\tt GALFIT} fit.}
\centering
\begin{tabular}{lcc}
%\hline\hline
%\\
Inner Disc \\
\hline
central surface brightness  & $\mu_{0,{\rm id}}$    & 17.5 \\
scalelength                         & $h_{\rm id}$             & 5.0 \\
ellipticity                             & $\epsilon_{\rm id}$   & 0.40 \\
position angle                     & PA$_{\rm id}$           & 90 \\
\hline
\\
Nucleus \\
\hline
effective surface brightness & $\mu_{e,{\rm n}}$    & 17.8 \\
effective radius                     & $r_{e,{\rm n}}$        & 1.0 \\
S\'ersic index                        & $n_{\rm n}$             & 1.1 \\
ellipticity                                & $\epsilon_{\rm n}$  & 0.30 \\
position angle                       & PA$_{\rm n}$          & 90 \\
\hline
\end{tabular}
\tablefoot{Luminosity parameters are in units of (F850LP) AB mag arcsec$^{-2}$. Spatial measurements are in units of arcseconds. Position angles are in degrees from North Eastwards. Note that the fit also includes a central point source. In addition, it also includes the major disc and the bar as found with the S$^4$G {\tt BUDDA} fit, i.e. the parameters for these components in that fit were kept fixed here.}
\end{table}

We now have a good understanding on the stellar components hosted by NGC 4371, in the context of their structural properties, such as geometrical properties. One interesting point already is the lack of any signature of a massive component that could be a classical, merger-built bulge. The components fitted by the photometric bulge model in the decomposition of the S$^4$G image are now understood to be the inner disc, the $10''$ ring, the $3''$ disc-like structure and the nucleus. All these components appear to be disc-related and not the outcome of a catastrophic event. This is not to say that they cannot be formed from accretion of external material, but a simpler explanation for their existence does not involve major mergers -- internal processes, such as evolution driven by disc instabilities such as the bar, are a more likely explanation.

\citet{ErwSagFab14} argue that NGC 4371 hosts a small classical, merger-built bulge. This claim is partly based on the evidence that the mean ellipticity in the inner $5''$ is smaller than that of the $10''$ ring and major disc, which indicates a rounder component. Consistently with this picture, Erwin et al. also show, using SINFONI data, that there is a significant increase in the importance of random stellar motion in this region. On the other hand, Figs. \ref{fig:hstprof}, \ref{fig:hstcolor} and \ref{fig:hstum} show that the $3''$ component is rather disc-like, as are the components making up what we are calling the galaxy nucleus, such as the dusty and the bluish ring-like structures clearly seen at the bottom panel of Fig. \ref{fig:hstcolor}.
%Figure \ref{fig:resid} even suggests that the $3''$ component could be a thick ring or a lens\footnote{If the $3''$ disc-like structure turns out indeed to be a ring, it opens up the exciting possibility that, together with the $10''$ ring, the two rings are at the inner and outer ILR, respectively.}.
In addition, our HST image fit yields a S\'ersic index of 1.1 for the nucleus component, indicating a disc-like structure (see Table \ref{tab:decomps2}; typical uncertainties in the measurement of S\'ersic indices are about 20\%), built from the major disc, since minor mergers are known to result in S\'ersic indices significantly larger than 1 \citep[see][but see also \citealt{QueEliTap15} and \citealt{ChrBroFis14}]{AguBalPel01}. We cannot say much about the point source component within a radius of about $0.15''$, however. Physically, this corresponds to about 12 pc, which means it could be a nuclear stellar cluster \citep[see][]{BokSarMcL04}. We will return to discussing whether NGC 4371 hosts a small classical bulge in Sect. \ref{sec:kin} below, where again our kinematical measurements will come to help.

Furthermore, from the results of our fits with {\tt BUDDA} and {\tt GALFIT} we also created what we call structural maps. These are simply masks, one for each component fitted, as well as for the $3''$ disc-like component and the $10''$ ring, that indicate in the MUSE field where each component dominates the emission of light from the galaxy, as seen with the S$^4$G and HST images. Thus, each spatial element is assigned to a single structural component, namely the one that dominates the light at that spatial element, above the other components that might as well contribute to part of the light in that spatial element. We therefore constructed structural maps for the central point source, the nucleus, the $3''$ disc-like component, the $10''$ ring, the inner disc, the bar and the major disc. These maps will be used below in order to enhance our understanding and interpretation of the results from the MUSE data cube regarding kinematics and stellar content.

\section{Stellar kinematics}
\label{sec:kin}

In this section, we first explain the steps taken to extract the kinematics information on NGC 4371 from the MUSE data cube, and then show the results thus obtained. We also connect these results with those from the previous section.

\subsection{Extraction of the kinematic maps}

\begin{figure}
\centering
\includegraphics[width=\hsize]{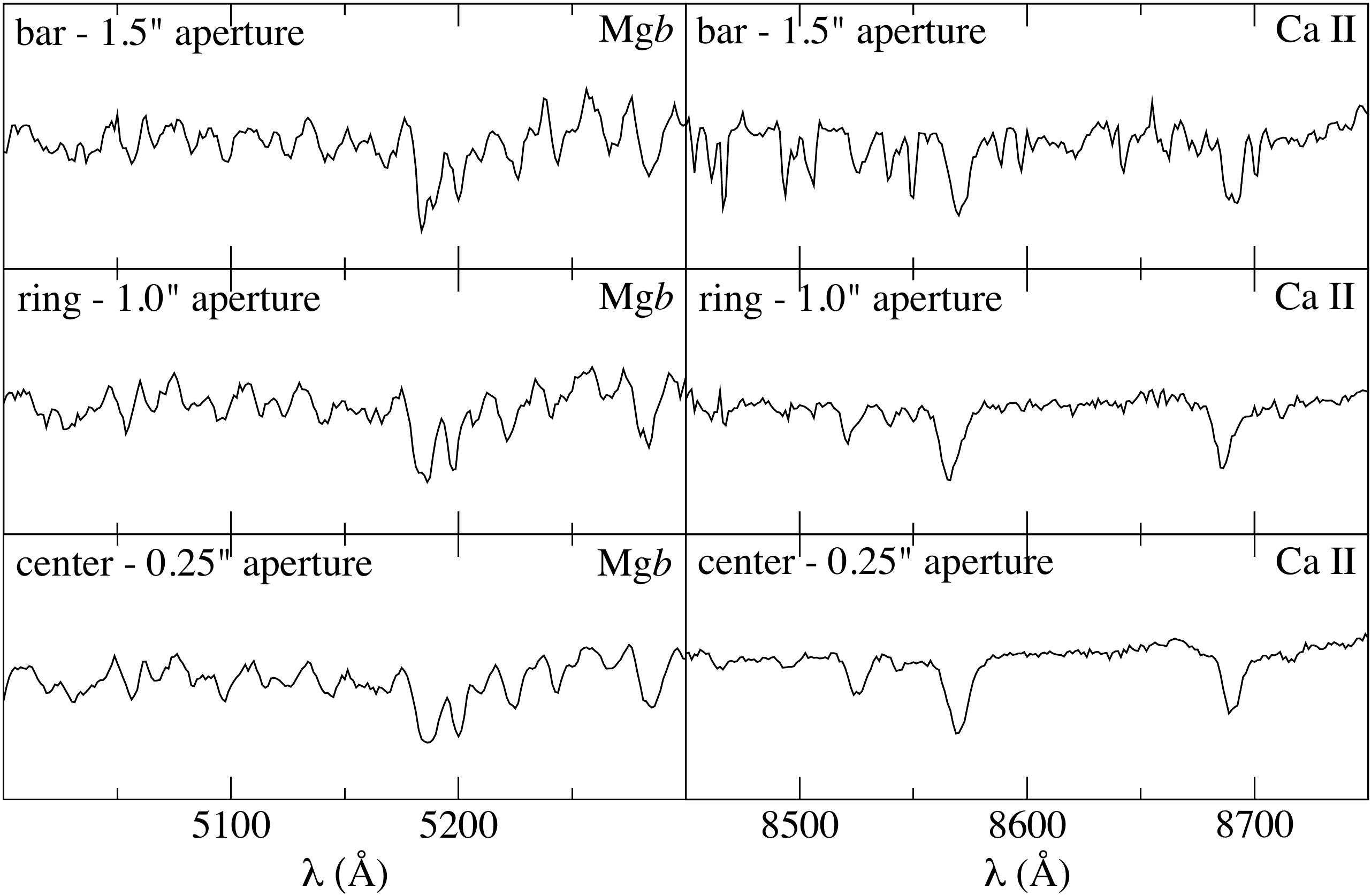}
\caption{Sample of spectra extracted from our MUSE data cube. Left-hand panels show the region around the Mg{\it b} feature, while right-hand panels show the region around around the CaII triplet. Top panels show the spectrum extracted from a region in the Northern part of the bar, $25''$ from the centre, with an aperture of $1.5''$, or 6 spaxels in diameter. Middle panels show the spectrum obtained from the East side of the $10''$ ring with an aperture of $1.0''$. Finally, bottom panels show the spectrum from the central spaxel.}
\label{fig:spectra}
\end{figure}

A high quality analysis of the stellar kinematics requires a minimum signal-to-noise ratio (S/N) in the spectra. To derive the radial velocity and the velocity dispersion that best represent the line of sight velocity distribution (LOSVD) at a given position in the galaxy, the corresponding minimum S/N is not a limiting condition. However, the derivation of the higher order Gauss-Hermite moments h$_3$ and h$_4$ of the LOSVD requires high S/N \citep[see][see also \citealt{GaddeS05}]{vanFra93}. Using a Voronoi binning scheme \citep{CapCop03}, we binned our data spatially to achieve a minimum S/N of $\approx100$ at each spatial element. Central pixels surpass this limit by far and therefore remain unbinned. In order to avoid contamination at each spatial element by spaxels with very weak signal, we also implemented a minimum S/N threshold of $\approx3$ for a spaxel to be considered. Therefore, the edges of the maps shown below from our MUSE data cube are not straight lines, since often pixels at these edges do not reach the minimum threshold in S/N. 

In order to illustrate the quality of the spectra used here, we show in Fig. \ref{fig:spectra} a sample of spectra extracted from our MUSE data cube, corresponding to different parts of the galaxy. As mentioned previously, our spectra cover the range from 4750\AA\, to 9300\AA\, but for clarity we show here only two spectral regions: one around the Mg{\it b} feature and other around the CaII triplet. While the central spectrum was extracted from a single spaxel, the spectrum from the East side of the $10''$ ring is a median combination of spectra within a $1''$ aperture, i.e. 4 spaxels in diameter. Likewise, the spectrum from the Northern part of the bar, at $25''$ from the centre, corresponds to an aperture of $1.5''$. These apertures correspond to the typical Voronoi spatial bins we used at these regions. One sees that this spatial binning is efficient in keeping the S/N at a suitable level throughout the field.

We used the pPXF -- penalized pixel fitting -- code developed by \cite{CapEms04} to extract the stellar kinematics, including higher order moments. The result is a LOSVD described by a Gauss-Hermite parametrisation \citep{Ger93,vanFra93}, with a measure of the radial velocity ($v$), velocity dispersion ($\sigma$) and higher order Gauss-Hermite moments (h$_3$ and h$_4$). The h$_3$ and h$_4$ moments indicate, respectively, asymmetric and symmetric deviations from a LOSVD that is a pure Gaussian. In other words, h$_3$ is basically a measure of the skewness of the LOSVD, whereas h$_4$ is essentially a parameter that quantifies the kurtosis of the LOSVD. They can be used to examine the orbital structure of the stellar system in question. For instance, circular motion results in h$_3$ values that are anti-correlated with $v$. And high values of h$_4$ suggest the superposition of structures with different LOSVDs \citep[see e.g.][and references therein]{BenSagGer94}. These values are determined through a fitting process where the routine combines a number of template spectra from a previously defined stellar spectra library to fit the galaxy spectrum of each element. Here we used a subset of the MILES single stellar population (SSP) model spectra \citep{VazSanFal10} with a mean resolution with FWHM of $\approx2.5$\AA\ \citep{FalSanVaz11} to match the spectral resolution in MUSE. The resolution of the model spectra is adjusted to the spectral resolution of the data before the fitting process. We cover the following range of ages and metallicities with the SSP template spectra: 0.1 Gyr to 17.8 Gyr, and $-0.40$ < [Z/H] $<+0.22$, respectively. Throughout this work we assume a Kroupa initial mass function \citep[IMF,][]{Kro01}. A wide wavelength range was employed in the fits, namely from 4750\AA\ to 8800\AA, which improves the S/N, as compared to narrow wavelength ranges.

Our MUSE spectra virtually reveal no emission lines, except for a weak [NII] emission at 6583\AA\ in the inner $1''$, where the nucleus and the central point source dominate. The [NII] equivalent width in the integrated spectrum obtained by combing all spectra within $1''$ is $\approx0.5$\AA, but H$\alpha$ is seen only in absorption. We will come back to this point in Sect. \ref{sec:discuss}, but we point out here that the lack of emission lines means that the results from our pPXF fits are straightforward and not complicated in this respect.

\subsection{The kinematics of the different structural components in NGC 4371}
\label{sec:kin2}

\subsubsection{Kinematic maps}

\begin{figure*}
\centering
\includegraphics[width=0.8\hsize]{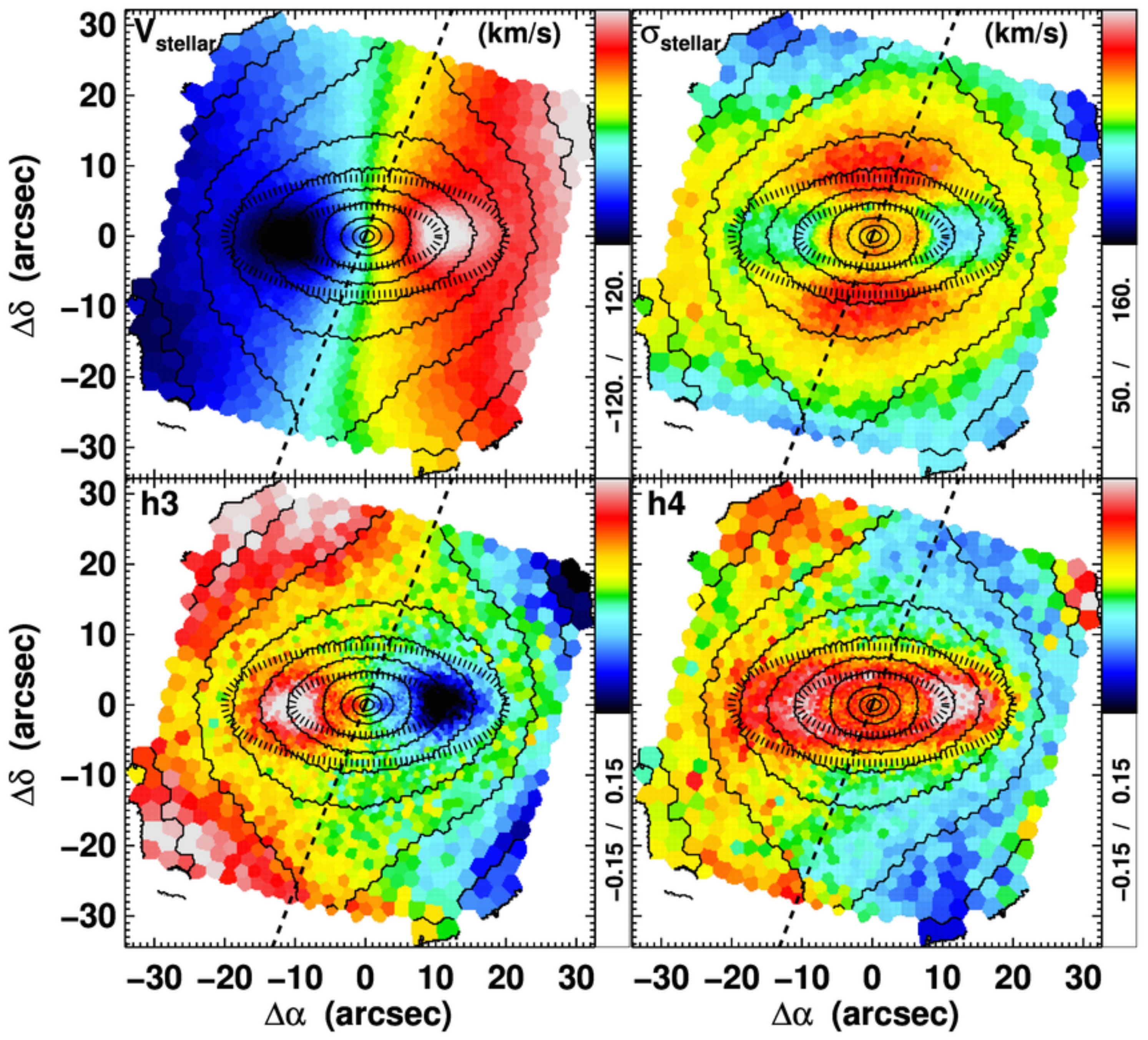}
\caption{Radial velocity, velocity dispersion, h$_3$ and h$_4$ maps for the stellar component in NGC 4371, as indicated. The colour bars on the side of each panel indicate the plotted range of the
parameter measured. For radial velocity and velocity dispersion these are given in km s$^{-1}$. The isophotes shown are derived from the MUSE cube reconstructed intensities and are equally spaced in steps of about 0.5 magnitudes. North is up, East to the left. The nearly vertical dashed line indicates the position of the bar major axis. The horizontal elongated markings indicate the outer boundary of the region dominated by the inner disc, at $\approx20''$ -- as derived through the ellipse fits -- and the position of the $10''$ ring.}
\label{fig:kin}
\end{figure*}

Figure~\ref{fig:kin} shows stellar kinematics maps of radial velocity $v$, velocity dispersion $\sigma$, and Gauss-Hermite moments h$_3$ and h$_4$. Intensity contours equally spaced in steps of about 0.5 magnitudes are indicated. NGC 4371 was also observed with SAURON \citep[see][]{BacCopMon01,deZBurEms02}, as part of the ATLAS3D survey \citep{CapEmsKra11}, covering the inner 30$\times$40 arcsec. Our MUSE data reveal a multitude of details not visible in that former data set. It is easy to contemplate the major improvement brought about with MUSE by the same PI and the MUSE consortium a decade later \citep{BacAccAdj10}.

The stellar velocity field exhibits an overall regular rotation pattern, with enhanced central regions hinting towards distinct kinematic components in those regions. This is confirmed in the other maps. The line of nodes is almost aligned perpendicular to the bar major axis and only shows a very slight twist in the region covered by the MUSE field, namely around $5''$ off the centre and in the bottom part of the field. The overall velocity seems to further rise towards the edges of the field, suggesting that we do not reach the peak in the rotation curve.

The stellar velocity dispersion is generally elevated within an almost circular region of radius of about $20''$, with several well-defined regions presenting varied patterns in terms of velocity dispersion. There are two large regions of higher velocity dispersion above and below the centre along the bar major axis, and other two large regions of lower velocity dispersion along the East-West direction. The very centre does not exhibit the highest velocity dispersion, but there seems to be four concentrated regions around the centre (and about a few arcseconds from it) with higher velocity dispersion, two of such regions aligned with the East-West direction, the other two aligned with the North-South direction.

The h$_3$ parameter strongly anti-correlates with the radial velocity, in particular in the regions of high  absolute radial velocity and low velocity dispersion, both in the inner regions, as well as at the edges of the field. The two regions showing the highest absolute values of h$_3$, at about $10''$ from the centre both at East and West, also show the highest h$_4$ values. There are also two well-defined, concentrated regions showing high absolute values of h$_3$ at a radius of about $1-3$ arcsec again both at East and West. These regions show counterparts with high radial velocities in the radial velocity map. The h$_4$ map also shows a region of high values that can be described as an elongated ring with semi-major axis reaching out to about $20''$ along the East-West direction. A similar elongated region around the centre, extending to $\sim5''$ also shows high h$_4$ values. Two of the four circumnuclear regions with high velocity dispersion show a hint of having relatively low h$_4$ values. These are the ones along the East-West direction. A similar relative drop in h$_4$ for the other two regions along the North-South direction is not so discernible. The large-scale structure in the h$_4$ map also displays asymmetric features on its edges.

\begin{figure*}
\centering
\includegraphics[width=0.8\hsize]{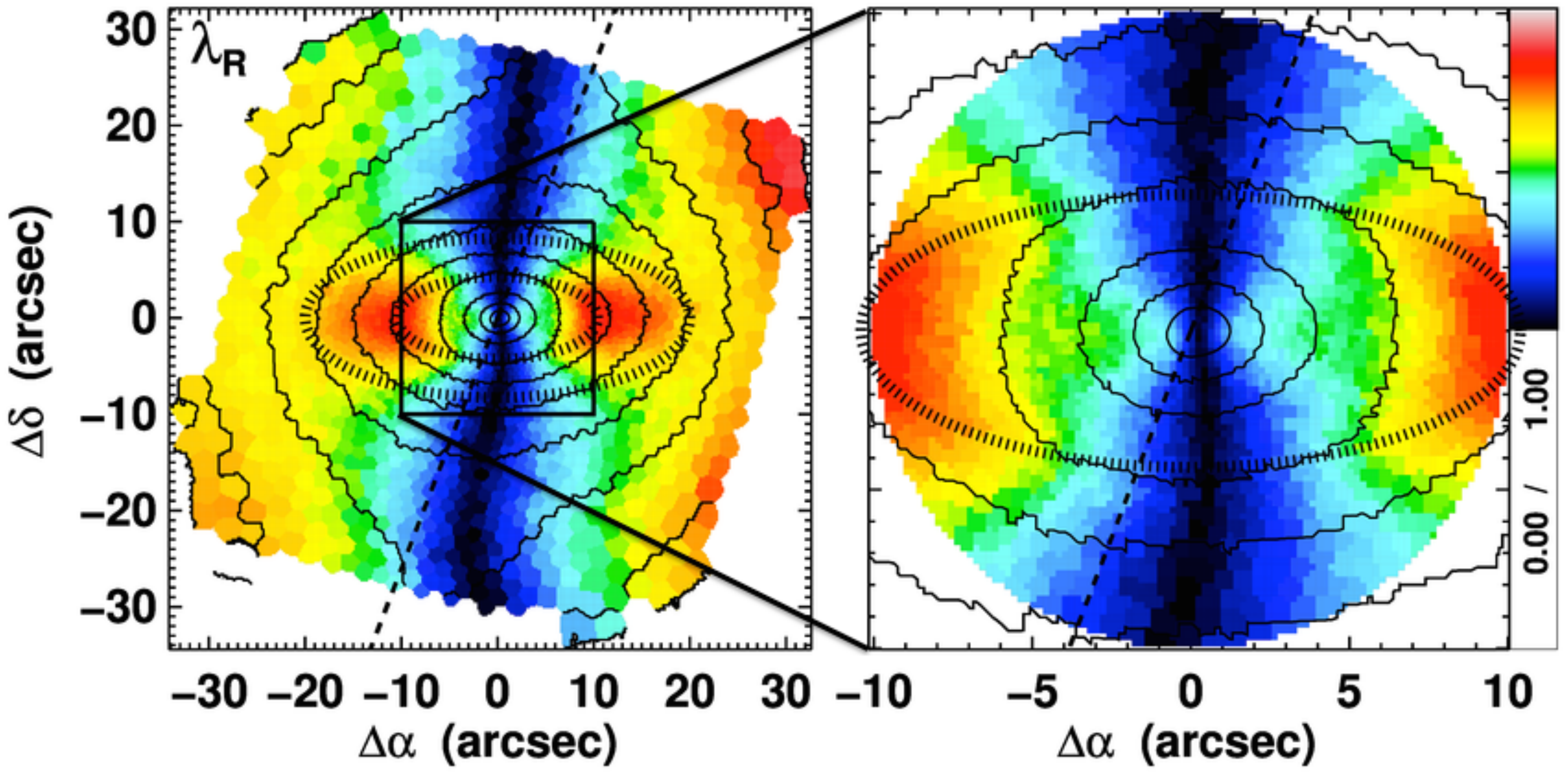}
\caption{Map of $\lambda_R$ for NGC 4371. The isophotes shown are derived from the MUSE cube reconstructed intensities and are equally spaced in steps of about 0.5 magnitudes. The right-hand panel focusses on the inner $10''$. North is up, East to the left. The nearly vertical dashed line indicates the position of the bar major axis. The horizontal elongated markings indicate the outer boundary of the region dominated by the inner disc, at $\approx20''$ -- as derived through the ellipse fits -- and the position of the $10''$ ring.}
\label{fig:lam}
\end{figure*}

Another important kinematical parameter to be considered here is $\lambda_R$. This parameter was introduced by \citet{EmsCapKra07} as a proxy to quantify the observed projected stellar angular momentum per unit mass. It is defined as:

\begin{equation} 
\lambda_R\equiv\frac{\left<R \, |v| \right>}{\left<R\sqrt{v^2+\sigma^2}\right>},
\end{equation}

\noindent where $R$ is the galactocentric radius. Figure \ref{fig:lam} shows the corresponding map. This map shows more clearly now the features that produce the pinching along the East-West direction in the inner $2-3$ arcsec. These features were already noticed above in the maps of radial velocity and h$_3$, and they correspond to relatively high angular momentum. This map also shows the high angular momentum component at $10-20$ arcsec from the centre along the East-West direction.

\subsubsection{The bar, box/peanut and barlens}

Now, how do all the features seen in these maps relate to the structural components discussed in Sect. \ref{sec:struct}? The high rotational support in the region dominated by the major disc is no surprise. The bar shows a remarkable region with high velocity dispersion. Because the galaxy is quite inclined ($i\approx60^\circ$), these measures of $\sigma$ have components of velocity dispersion both at the plane of the disc and at the direction perpendicular to it. Could these enhancements in $\sigma$ be caused by a box/peanut? To try answering this question we compare our maps with those recently published by \citet{IanAth15}. Iannuzzi \& Athanassoula have produced maps similar to ours from simulations of barred galaxies at different evolutionary stages, before and after the formation of the box/peanut in the bar. Their figures 29 and 31 show their simulation GTR101 at a very similar projection as how we see NGC 4371. Their velocity dispersion maps show enhancements along the bar major axis that very much resemble our results for NGC 4371, particularly so after the formation of the box/peanut. Another signature of box/peanuts pointed out by \citet{DebCarMay05}, and corroborated by \citet[][see also \citealt{MenCorDeb08}]{IanAth15}, are minima in h$_4$ seen along the bar major axis at each side of the centre.  Our h$_4$ measures along the bar major axis suggest the presence of such minima, particularly at the bottom of the field. Nevertheless, we admit that this evidence is only suggestive. \citet{ErwDeb13} do not find evidence of a box/peanut in NGC 4371 but they cannot rule out its existence. Overall, it seems plausible that NGC 4371 has a box/peanut. Such structures are known to reduce the strength of bars in inducing secular evolution processes and might significantly influence the evolutionary processes in a galaxy, such as gas flows \citep[see][]{FraAthBos15}.

As mentioned above, recent work have put forward arguments suggesting that box/peanuts have a counterpart in the plane of the disc, which are barlenses \citep{LauSalAth14,AthLauSal14,Ath15,LauSal15}. We do not find evidence for a barlens in NGC 4371 in the kinematic maps. In the structural analysis done in Sect. \ref{sec:struct} the only structural component found that could be a barlens is what we call the inner disc, which dominates the emission of light from the galaxy from a radius of about $10''$ to about $20''$. Figures \ref{fig:kin} and \ref{fig:lam} show clearly that this component is strongly rotationally supported. It is not clear yet whether this would be consistent with a barlens, but it is certainly consistent with an inner disc indeed. Therefore, if NGC 4371 has a box/peanut, it could be that it is not strong enough to produce an evident signature of its disc plane counterpart, the barlens, in our data. In addition, projection effects and/or the presence of the other conspicuous inner components in the same region -- such as the inner disc, in fact -- could mask out signatures of the barlens.

\subsubsection{The inner disc, $10''$ ring and inner structures}

The $10''$ ring is just at the inner edge of the inner disc and cannot be distinguished from it in Figs. \ref{fig:kin} and \ref{fig:lam}, except perhaps in the h$_4$ map. The fact that this separation is not straightforward hints at a connected formation history. At a radius of about $10''$ one sees a peak in  h$_4$, higher than the values seen between about $10''$ and $20''$, i.e. through the whole extent of the inner disc. The peak in h$_4$ can be explained if the inner disc and the $10''$ ring have different LOSVDs.

The $3''$ disc-like component can also be seen in the kinematic maps. It is a fast rotating central component, seen as the two concentrated peaks, near the centre, of absolute radial velocity and h$_3$, and $\lambda_R$. The major disc, the inner disc, the $10''$ ring and the $3''$ disc-like structure all show a strong anti-correlation between radial velocity and h$_3$, consistent with circular motion. Therefore, all such structural components, found through the image decompositions, have their existence corroborated and better understood through the kinematic maps. The central point source and the components that we have collectively called the galaxy nucleus dominate the central $1''$.

Now we can come back to the question on whether NGC 4371 hosts a small classical bulge, within $5''$ from the centre, as suggested by \citet{ErwSagFab14}. Our maps show clearly that the $3''$ disc-like component is strongly supported by rotation and circular motion, which again argues against the presence of a classical bulge that extends to a few arcseconds. [Unless, somehow, a component with relatively high angular momentum results from the bulge formation processes, despite their violent nature. Another possibility is that the small classical bulge did not originally have significant angular momentum, but that it was spun-up by the bar \citep[see][]{SahMarGer12}.] Furthermore, we note that, to the spatial resolution limit that we can achieve with these observations (i.e. $\lesssim1''$), the central values of velocity dispersion are below those along the inner part of the bar major axis, i.e. there is no discernible kinematically hotter component in our central kinematic bins. A caveat here, however, is that, if the light from the central bins is biased towards a relatively younger population that is kinematically colder -- even though, as we will see shortly, the stellar population content in the centre is predominantly very old -- this could mask an older, kinematically hotter component. The measurements presented in Fig. 6 of \citet{ErwSagFab14} were performed in the near-infrared and with adaptive optics, using SINFONI at the VLT, and therefore reach a spatial resolution much better than our own measurements (i.e. $\approx0.1''$). From a kinematical perspective, we cannot rule out that the nucleus was built through mergers, but we remind the reader that our fits to the HST image of NGC 4371 result in a nucleus with an exponential, disc-like light profile. In addition, there is a peak in ellipticity at a radius of $\approx0.4''$, where the nucleus dominate. There is also very little we can say about the central point source here.

\section{Stellar population content}
\label{sec:pops}

As in the previous section, here we will also first focus on the methodology employed, in this case to study the properties of the different stellar populations in NGC 4371, and then show the results thus obtained. We will connect these results with the results from our structural and kinematical analyses above, and, in particular, make use of the structural maps described in the last paragraph of Sect. \ref{sec:struct}.

\subsection{Extraction of the stellar age and stellar metallicity maps}
\label{sec:ages1}

In this paper, we have chosen to perform a full spectral fitting using the code {\tt STECKMAP} (STellar Content and Kinematics via Maximum A
Posteriori likelihood, see \citealt{OcvPicLan06a,OcvPicLan06b}). The reconstruction of the stellar age distribution and the age-metallicity relation
is non-parametric, i.e. no specific shape for the  star formation history is assumed. The problem of obtaining this physical information
from the observations is an ill-conditioned problem \citep{OcvPicLan06a}, where small fluctuations in the data can result in large
variations in the solution. To deal with this problem, {\tt STECKMAP} makes a statistical regularisation with the {\it prior} in such a way that solutions that
change smoothly are more likely to result. This is the only {\it a priori} condition. The smoothness parameter can be set by generalised
cross-validation of the possible solutions, accounting for the level of noise in the data, in order to avoid over-interpretation of the results \citep[see][for details]{OcvPicLan06a}.

The function to minimise is defined as:

\begin{equation}
Q_{\mu} = \chi^2 (S(x,Z,g))+ P_{\mu}(x,Z,g),
\end{equation}

\noindent which is a penalised $\chi^2$, where $S$ represents the synthetic spectrum, $x$ the flux distribution, $Z$ the metallicity
distribution, and $g$
the broadening function. 
The penalization $P_{\mu}$ can be written as $P_{\mu} (x,Z,g) = \mu_x P(x) + \mu_Z P(Z) + \mu_v P(g)$, where the function $P$
gives high values for solutions with strong oscillations (flux or metallicity changing rapidly with time) and small values for
smoothly varying solutions. Adding the penalization $P$ to the objective function is exactly the same as injecting an {\it a priori}
probability density distribution to the solution, as $f_{\rm prior}(x)=\exp(-\mu_x P(x))$, where $P$ is a quadratic function of the unknown.

We use again the MILES SSP spectral models \citep{VazSanFal10}, spanning an age range from 6.3$\times$10$^7$ to
1.7$\times$10$^{10}$ years, divided in 30 logarithmic age bins. The metallicity ranges from $[Z/H]= -1.3$ to $[Z/H]=+0.2$.

Although {\tt STECKMAP} can be used to obtain the broadening of the lines, here we use the values obtained in the previous section, calculated with pPXF, and thus fix the broadening function during the fit. The reason for this choice comes from the existing degeneracy between metallicity and $\sigma$ \citep{KolPruDer08,SanOcvGib11}, which biases the mean-weighted metallicity if both parameters are fitted at the same time.
We also do not fit the continuum, to avoid dealing with possible flux calibration errors or dust extinction. Instead, we
multiply the model by a smooth non-parametric transmission curve. This curve is obtained by interpolating a spline function between 30 nodes
spread uniformly along the wavelength range.  The choice of fitting or not the continuum may have a strong influence in the
derived parameters when the S/N of the spectrum is not very high. For spectra with S/N $>30$ per \AA\, the results obtained are independent on rectifying or not the continuum \citep{SanOcvGib11}.

\begin{figure}
\centering
\includegraphics[width=\hsize]{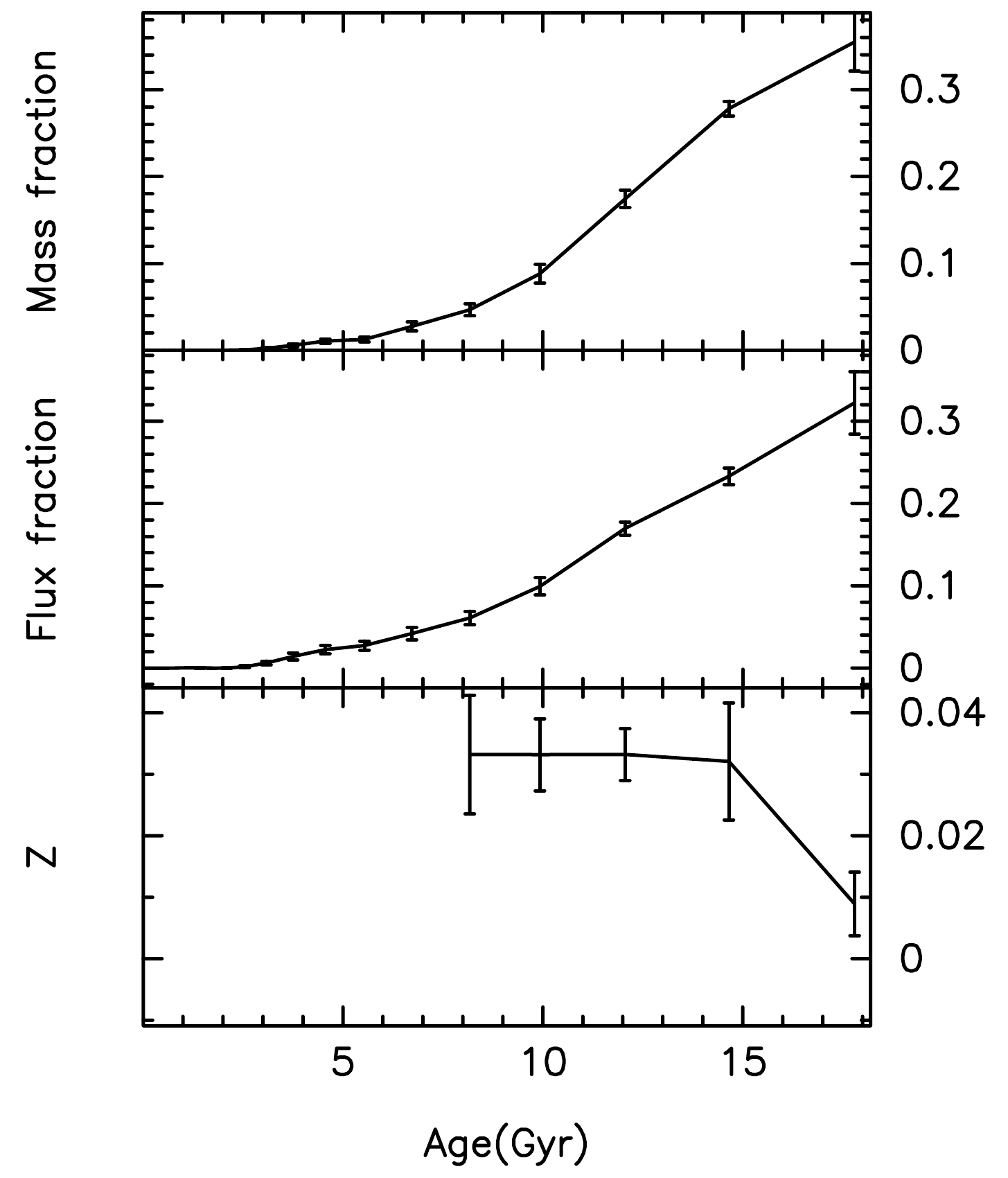}
\caption{Mass (top panel) and flux (middle panel) fractions as a function of age, and 
the age-metallicity relation (bottom panel), obtained for the central spectrum of NGC 4371. 
Metallicities are plotted only for flux fractions larger than 5\%.}
\label{fig:SFH_stecmap}
\end{figure}

We ran {\tt STECKMAP} on each of the binned spectra, and obtained star formation histories and age-metallicity relations for all of
them. Figure \ref{fig:SFH_stecmap} shows an example of a typical {\tt STECKMAP} output, showing the flux, mass and metallicity distributions as a function of age.

We can obtain mean values of age and metallicity per spectrum weighting with either the flux or the mass. Flux-weighted values are biased towards younger populations, as young stars are brighter in the considered wavelength range. Mass-weighted values may be, on the other hand, much more uncertain, especially when a large fraction of old, low-luminosity stars are present. To better understand how uncertain are our mean stellar age estimates at each spatial resolution element -- which will be discussed at length below -- we have performed 100 Monte Carlo realisations using all binned spectra. We find that the error in luminosity-weighted age is anti-correlated with the mean age in a logarithmic scale. Considering luminosity-weighted estimates, for a mean stellar age of 10 Gyr the median of the error distribution is 0.035 dex, while for a mean stellar age of 7 Gyr the median error is 0.05 dex. Therefore, for these mean ages, the uncertainty in the mean stellar age is $\sim0.8$ Gyr. This anti-correlation is not present, however, when we consider mass-weighted estimates, which have a median error of 0.014 dex. One should note that the Monte Carlo realisations only address uncertainties of a Poissonian nature. Uncertainties arising from the spectral models employed are not included. In addition, our age measures may have as well systematic effects towards higher values, as we have measures reaching ages slightly older than the currently accepted age of the universe, in particular when one considers mass-weighted estimates. In an attempt to neutralise this effect, our main conclusions below are based on the luminosity-weighted estimates, which are systematically lower than the mass-weighted estimates. In addition, we consider the lowest age values whenever there is a wide enough distribution of mean stellar ages.

\subsection{Spatial distribution of the different stellar populations}

\begin{figure*}
\centering
\includegraphics[width=0.8\hsize]{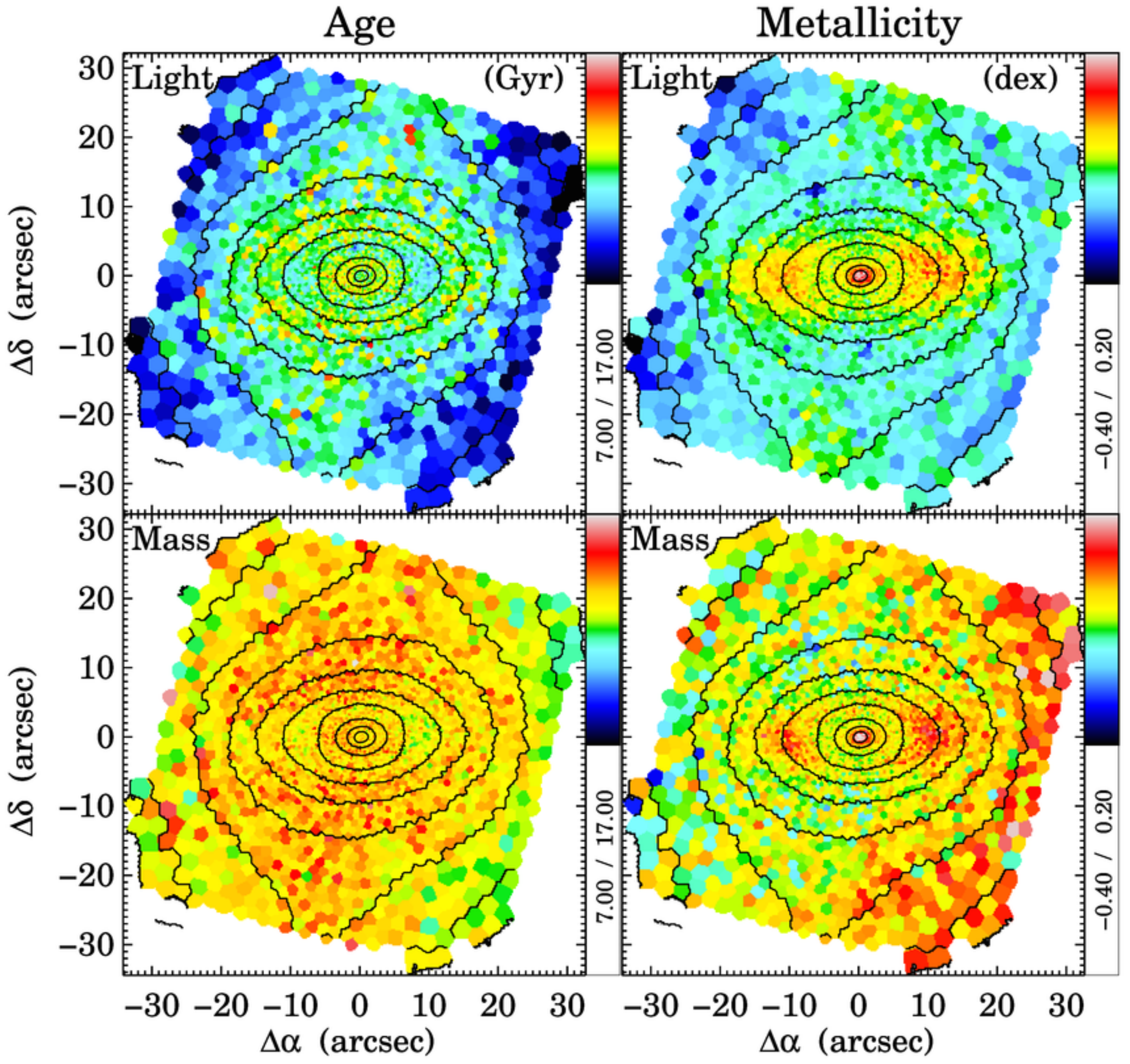}
\caption{Maps for NGC 4371 of mean stellar age (left panels) and stellar metallicity (right panels) weighted by both luminosity (top panels) and mass (bottom panels). The age maps have units of Giga-years, and metallicities are given in the spectroscopic notation (logarithmic scale normalised to the solar value).}
\label{fig:stellarpop_maps}
\end{figure*}

Figure~\ref{fig:stellarpop_maps} shows the luminosity- and mass-weighted maps of age and metallicity for NGC 4371, derived using the relations above. The mass-weighted mean stellar age map shows that the vast majority of stars in the galaxy are old, with ages above 10 Gyr. The major disc is seen to show stars that are on average somewhat younger than the rest of the galaxy, but still old. Similarly less old stars on average appear also just inwards of the West side of the $10''$ ring. Bins containing the oldest stars are seen preferentially in parts of the bar and the inner disc. The luminosity-weighted map shows more structure. It shows again less old stars in the disc and in parts just inwards of the $10''$ ring (i.e. in the inner part of the inner disc), especially its West side. The bar shows a somewhat intermediate age population, with mean ages between those of the major disc and the remaining central components. Still, even in the luminosity-weighted map, almost all bins inwards of the region dominated by the major disc show mean stellar ages above 10 Gyr.

From the luminosity-weighted metallicity map, it can be seen that the disc shows a very homogeneous metallicity distribution. In addition, the outer parts of the bar show a slightly elevated metallicity, while the central regions show a remarkable peak in metallicity. Finally, the $10''$ ring shows as well elevated metallicity, in comparison to the outer components. The mass-weighted map shows less features, but also indicates a very strong peak in the central metallicity and the elevated metallicity of the $10''$ ring. \looseness -1

\begin{figure*}
\centering
\includegraphics[width=0.9\hsize]{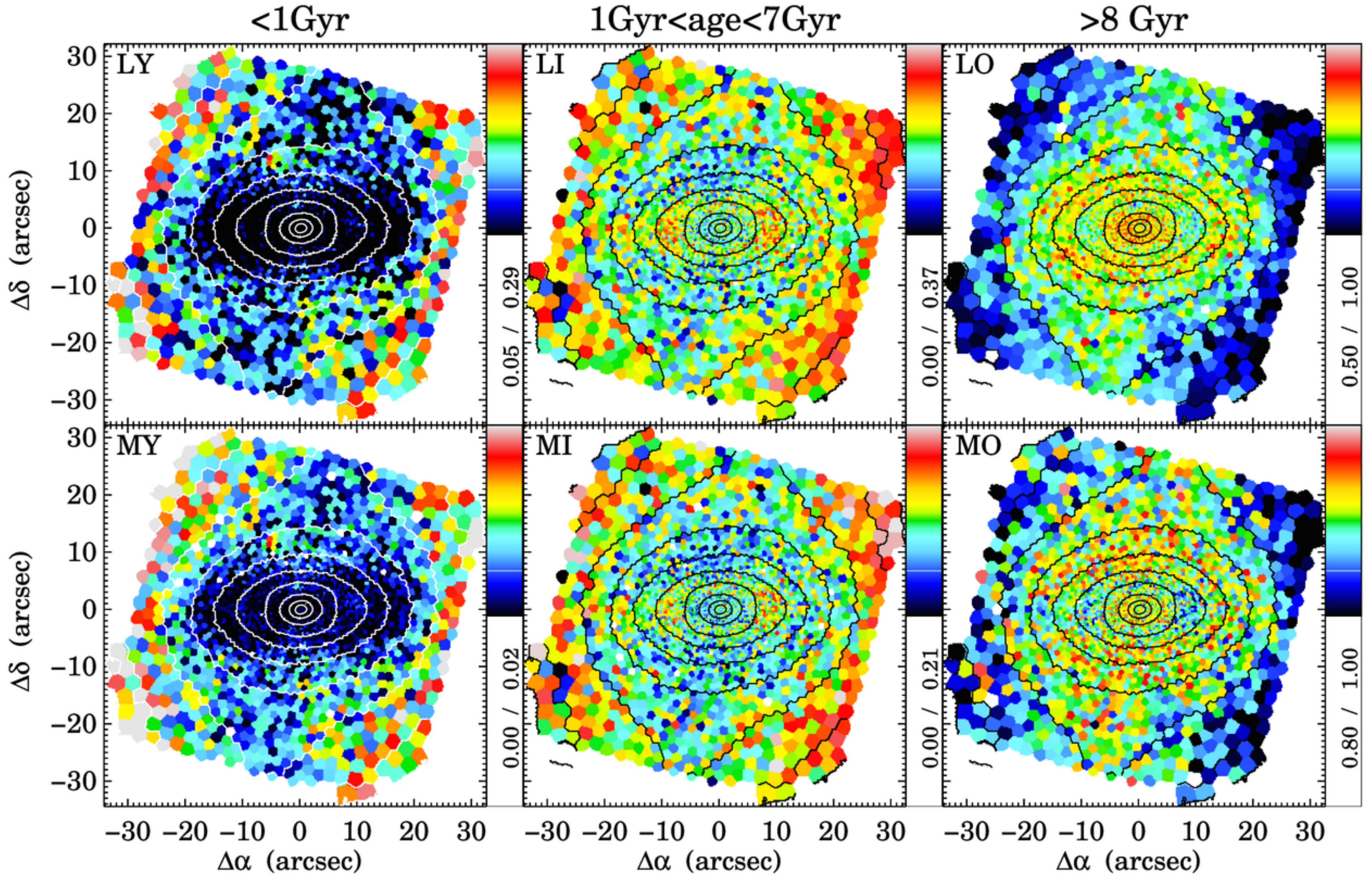}
\caption{Top panels: luminosity-weighted maps of stellar populations separated by mean stellar age, as indicated (Young; Intermediate; Old). The colour-coded scale, indicated by the colour bar at the right-hand side of each panel, corresponds to the fraction of the stellar population (young, intermediate age or old) at each spatial element. Bottom panels: corresponding mass-weighted maps. Note that the scale changes for each panel, in order to show the corresponding relevant features. The indicated limits of the colour bars, however, show that most of the mass and light are in (or come from) stars older than 8 Gyr.}
\label{fig:SP_sub_var}
\end{figure*}

In Fig. \ref{fig:SP_sub_var}, we show luminosity-weighted and mass-weighted maps that represent the location of three different stellar populations in NGC 4371, according to their ages: young (age $<1$ Gyr), intermediate ($1<{\rm age}<7$ Gyr), and old (age $>8$ Gyr). They roughly correspond respectively to a formation redshift of around $z\lesssim0.08$, $0.08\lesssim z\lesssim0.8$ and $z\gtrsim1$ (although we stress again here that the correspondence between mean stellar age and formation redshift is not trivial, given the uncertainties in the age estimates). The colour-coded scale corresponds to the fraction of the corresponding stellar population (young, intermediate age or old) at each spatial element. Note that the scale changes for each panel, in order to show the corresponding relevant features. The scale ranges are indicated below the colour bars on the right-hand side of each panel. One can see that most stars in the central arcminute squared of NGC 4371 are older than 8 Gyr. Figure \ref{fig:SP_sub_var} shows that basically one only finds stars formed at redshifts $z\lesssim0.08$ at the outskirts of the field (i.e. in the major disc), and to a much lesser extent in the bar. Stars formed between about 1 and 7 Gyr ago are found preferentially in the major disc and in the portion of the inner disc inwards of the $10''$ ring. To a lesser extent again, some such stars can be found in the bar. These younger stars, however, represent an insignificant fraction of the stellar content in this region of the galaxy probed by our MUSE observations. One important point to make here is that the mean stellar age in all these bins in which stars younger than 7 Gyr are found is actually above 8 Gyr, as demonstrated in Fig. \ref{fig:stellarpop_maps}. This will be further illustrated below. The outer part of the inner disc and the inner $5''$ show the highest densities of stars formed more than 8 Gyr ago.

\begin{figure}[h!]
\centering
\includegraphics[width=\hsize]{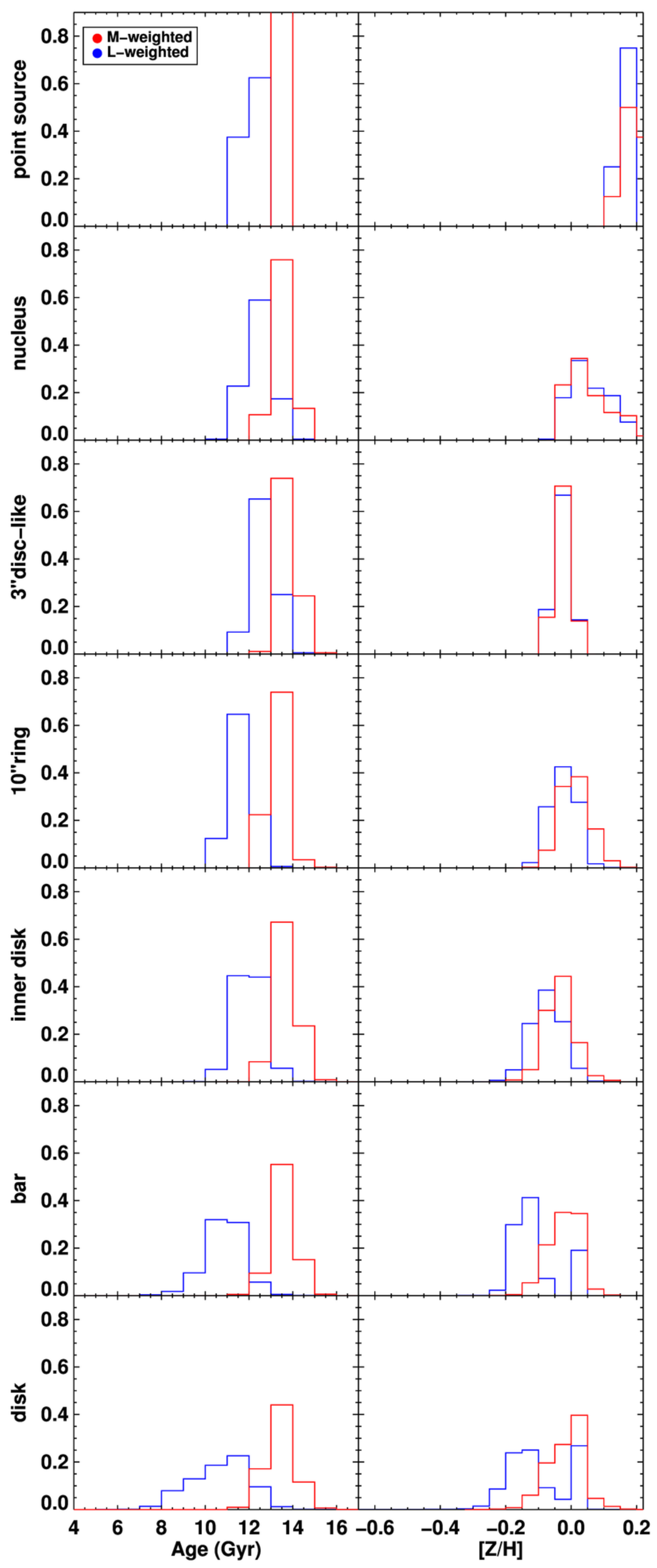}
\caption{Normalised distributions of the mean stellar age and metallicity per spatial bin, separated by structural component, as indicated. The mass-weighted distributions are plotted in red, whereas the luminosity-weighted distributions are plotted in blue.}
\label{fig:distr}
\end{figure}

There are other powerful ways of visualising the information content in such maps that may disclose important clues in a more straightforward way. One of such ways is presented in Fig. \ref{fig:distr}, which shows the normalised distributions of the mean stellar age and metallicity of the spatial bins in each structural component. These were derived through a combined use of the structural maps constructed in Sect. \ref{sec:struct} and the age and metallicity maps produced in this section (Fig. \ref{fig:stellarpop_maps}). These distributions corroborate our previous assessments and also reveal further aspects of the stellar content in NGC 4371. The luminosity-weighted metallicity distributions appear to be bimodal for the major disc and bar components only. Their {\em mass}-weighted metallicity distributions show a peak that roughly coincides with the high metallicity peak of their luminosity-weighted distributions, but they also show a tail to lower metallicities. Another interesting feature clearly seen in the luminosity-weighted age distributions is that all components closer to the centre than the $10''$ ring have on average an older population of stars than the remaining of the galaxy, particularly the $3''$ disc-like component. The mass-weighted distributions, however, show much subtle differences: they show that the whole stellar content of the galaxy is almost entirely very old.

\begin{figure}
\centering
\includegraphics[width=\hsize]{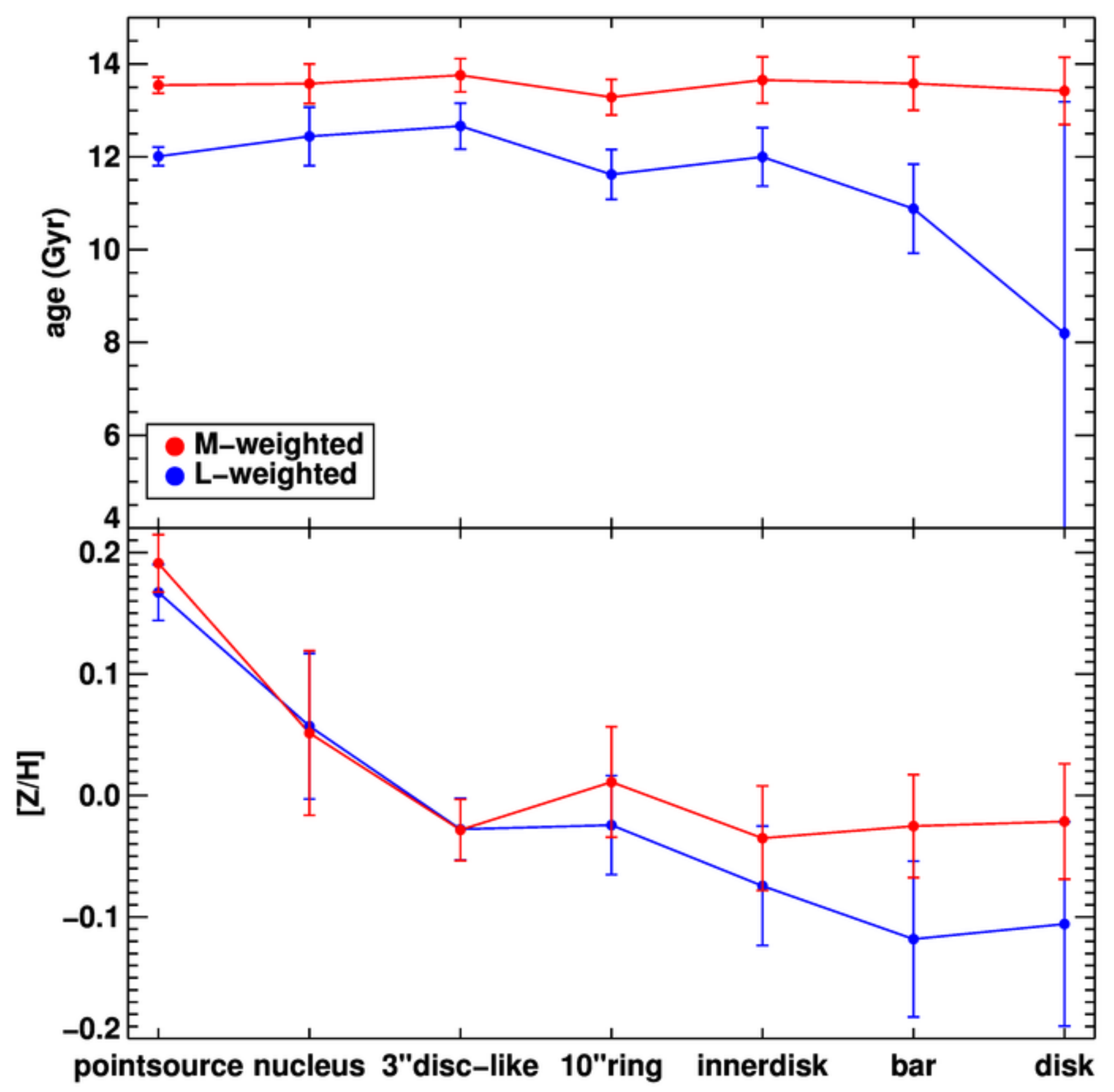}
\caption{Mass-weighted and luminosity-weighted mean values of stellar age and metallicity for all structural components in NGC 4371. The structural components are arranged along the x-axis such that this axis can be thought of as the galactocentric distance. I.e., the point source component is at the centre of the galaxy, and to the right-hand side of it we arranged the other components that dominate the emission from the galaxy at successively larger distances. The error bars are an outlier-resistant measure of the dispersion about the mean, analogous to the standard deviation.}
\label{fig:struct_pops}
\end{figure}

Finally, we present in Fig. \ref{fig:struct_pops} plots of the mass-weighted and luminosity-weighted mean values of stellar age and metallicity for all structural components in NGC 4371. Since the components are arranged along the x-axis in a way that from left to right each component successively dominates at larger galactocentric distances, these plots can be thought of as stellar age and metallicity radial profiles. Again, this figure repeats some of the information already discussed above, but this visualisation format allows a more direct comparison of the mean values of each component. In addition, it shows in a straightforward fashion how these values vary from the centre to the outskirts of the MUSE field. \looseness -2

The mean stellar age profiles show how homogeneously old is the stellar population in the inner arcminute of NGC 4371 (particularly so in the mass-weighted results), with the indication that younger stars are preferentially found in the outer part of the field, in the major disc. [Similar results are found by \citet{SeiCacRui15} for NGC 5701, another early-type barred galaxy.] A mild drop in the mean stellar age is also seen at the $10''$ ring. The mean stellar metallicity profiles clearly show the pronounced peak in metallicity at the central components. Again, the $10''$ ring produces a bump in the curves, this time indicating a higher metal content than what expected from the global trends of the profiles, i.e. as compared to a metal content that grows smoothly towards the centre. These profiles show essentially that beyond the inner few central arcseconds the mean metallicity does not vary much, whereas it starts increasing in the central arcsecond until it reaches a maximum at the very centre. The luminosity-weighted profiles in Fig. \ref{fig:struct_pops} show a somewhat different behaviour when compared to the mass-weighted profiles, possibly indicating different properties between a more luminous and {\em relatively} younger stellar population and the old bulk of the stellar content. These differences, along with all the results presented in this section and in Sects. \ref{sec:struct} and \ref{sec:kin} will be interpreted and discussed in the next section.

\section{Discussion}
\label{sec:discuss}

\subsection{NGC 4371: a fossil record of the oldest bars}

Being it a disc galaxy, and one that hosts a large-scale bar, one expects that internal secular evolution processes have played an important role in the evolution of NGC 4371. The {\tt BUDDA} fit (see Table \ref{tab:decomps1}) indicates that the bar-to-total luminosity ratio in the galaxy (Bar/T $=0.077$) is typical of the values found for other massive galaxies. \citet{Gad11} finds that the typical Bar/T is about 0.1 in a sample of nearly 300 massive barred galaxies at $z\approx0.05$ \citep[see also][]{WeiJogKho09}. Interestingly, the luminosity profile of the bar in NGC 4371 is very flat. This is indicated by its low S\'ersic index, namely $n_{\rm bar}=0.2$. \citet{KimSheGad15} have conjectured that recently formed bars have exponential luminosity profiles, i.e. with a S\'ersic index $\approx1$, but, as bars evolve, their luminosity profiles become flatter, which translates to lower values of $n_{\rm bar}$. The flattest bars in the work of \citet{KimSheGad15}, with 144 nearby barred galaxies from the S$^4$G, have a S\'ersic index similar to that of the bar in NGC 4371. This suggests thus that the bar in NGC 4371 formed at the epoch were the first long-standing bars were being formed. Being an evolved bar, it makes it more likely that a box/peanut has formed, as such structures are seen in simulations to form $\approx1-2$ Gyr after the formation of the bar \citep[see e.g.][]{MarShlHel06}.

As discussed in the Introduction, one of the main processes induced by large-scale stellar bars in disc galaxies is the inflow of gas towards the central regions. This inflow of gas is halted near the bar ILR and originates nuclear rings and inner discs (also known as disc-like bulges) with recently formed stars. We have suggested above that because the inner disc and the $10''$ ring in NGC 4371 are hardly distinguishable in some of the kinematic maps, they could have a connected formation history. Therefore, the most natural explanation for their origin is that of bar-driven gas inflows. Figure \ref{fig:distr} shows that there is no spatial bin in the inner disc or the $10''$ ring with a mean stellar age below 10 Gyr. Thus, the last significant gas inflow induced by the bar must have happened 10 Gyr ago, and then it is not only the case that the bar in NGC 4371 is populated by old stars but the bar itself, as a structure, is old. {\em In fact, the mean stellar ages of the inner disc and $10''$ ring imply that the bar was already in place at least 10 Gyr ago, i.e. at a redshift of about 1.8}. Considering the results from the analysis of the uncertainties in our stellar age estimates, described at the end of Sect. \ref{sec:ages1}, the formation redshift of the bar is then most likely between $z=1.4$ and $z=2.3$.

In the Introduction, we mentioned the study of \citet{SimMelLin14} on the fraction of barred galaxies at high redshifts. In that work, the highest redshift galaxy found hosting a bar is at $z=1.97$, similar to the formation redshift we derive for NGC 4371. The latter is then a fossil record of those bars, such as the ones found by Simmons et al., that formed at early cosmic epochs. Interestingly, the bar in NGC 4371 thus testifies to the robustness of bars.

Alternatively, the inner disc and the $10''$ ring could have formed via a yet unfamiliar mechanism \citep[see discussion in][see also \citealt{EliGonBal11}]{ComSalLau14}, which could mean that one cannot constrain the formation redshift of the bar using their mean stellar ages. One possibility is that they were formed by a bar that has dissolved before the formation of the current one. \citet{KraBouMar12} show simulations in which bars that form at redshifts larger than 1 dissolve, and a long-standing bar forms afterwards. However, it is still unclear whether such primordial bars would be able to form long-standing inner discs and nuclear rings. It is thus unclear if the $10''$ ring would survive the dissolution of the primordial bar and the formation and evolution of the bar currently observed. On the other hand, \citet{ComKnaBec10} report a fraction of $19\pm4$ per cent of nuclear rings found in unbarred galaxies. Nevertheless, the $10''$ ring in NGC 4371 is similar to the other 78 nuclear rings in the barred galaxies studied by \citet{ComKnaBec10}, in that the ratio between the ring radius and the bar length in NGC 4371 is very typical, consistent with a picture in which the nuclear ring is formed by the bar. Overall, it is thus unlikely that the inner disc and $10''$ ring in NGC 4371 were not formed by the bar.

\subsection{The onset of bar-driven secular evolution in cosmic history}

The fact that NGC 4371 is a massive galaxy with an old bar is consistent with the results from \citet{SheElmElm08,SheMelElm12}, which indicate that the more massive disc galaxies form their bars first. For $M_\star\geq10^{11}\rm{M}_\odot$, \citet{SheElmElm08} find that about half of the disc galaxies are barred already at $z\sim0.8$. One interesting outcome of the present study is that we can provide an estimate of the formation redshift for the bar in NGC 4371, and, since we know the galaxy stellar mass, we can go a step further and use this information as a benchmark to set how the bar formation epoch evolves as a function of galaxy stellar mass. Therefore, with a bar formation redshift of $z\approx1.8$ and a stellar mass $M_\star=10^{10.8}\rm{M}_\odot$, NGC 4371 sets this benchmark and establishes that more massive galaxies have formed their bars at even higher redshifts.

It is interesting to consider then, that the disc in NGC 4371 already had -- at these relatively high redshifts -- the structural and dynamical conditions favourable to the development of a long-standing bar. The internal bar-mode disc instability can happen spontaneously to any disc galaxy under the right conditions -- which are, nevertheless, complex and challenging to spot \citep[see e.g. discussion in][]{Ath08b,SanGad13} -- but might also be catalysed by the perturbation of a companion galaxy. A number of studies have been dedicated to understand the effects of the environment on bar formation \citep[e.g.][]{AguMenCor09,LiGadMao09,BarJabDes09,MenSanAgu10,MenSanAgu12,SkiMasNic12,LinCerLi14}. While it is interesting to consider the reasonable possibility that the bar in NGC 4371 is the result of an interaction, since the galaxy is in a dense environment, this is a difficult claim to defend. In any case, there is no observational evidence to date that bars catalysed by encounters are in any way different from bars formed spontaneously.

\subsection{Environmental effects: the lack of gas and recent star formation}

The vast majority of the stars in the central region of NGC 4371 have ages above $\approx8$ Gyr (see Fig. \ref{fig:stellarpop_maps}), and thus were formed at redshifts larger than about 1. In addition, there are no signs of a significant recent interaction throughout the galaxy. The galaxy thus appears to have gone through no significant accretion of cold gas to its central region for the past 8 Gyr.

As mentioned above, the only emission line we find in the spectra is a weak [NII] line confined within the central $1''$ and with equivalent width of $\approx0.5$\AA. The [NII] flux could originate from stellar photoionization powered by young high-mass OB stars, or from other physical processes, such as shock heating or photoionization with a power-law spectrum \citep[see e.g. discussion in][]{JamShaKna04}. The central supermassive black hole in NGC 4371 -- with a mass determined via dynamical modelling (P. Erwin et al., in prep.) -- is currently not accreting \citep{SimStodeF07}, so we will assume that all [NII] emission comes from recent star formation activity. Even so, the measured equivalent width indicates a very low star formation activity currently at the centre of NGC 4371 \citep[see Fig. 3 in][]{Ken98}. \citet{NagFalWil05} and \citet{CapKhaAxo09} find as well at best very low radio emission from the nucleus of NGC 4371, also consistent with low star formation or AGN activity and a low amount of cold gas. As mentioned at the end of Sect. \ref{sec:4371}, other studies have found very little atomic and molecular gas content throughout the galaxy.

Being it a member of the Virgo cluster, known effects at such dense environments could have been responsible for the lack of significant star formation and gas content in the central region of the galaxy in the past 8 Gyr. As we have seen, tidal interactions and mergers with other galaxies do not appear to have played a major role. Ram pressure stripping, however, could have removed gas from the galaxy, via the interaction between the galaxy cold inter-stellar gas and the hot intra-cluster gas \citep{GunGot72}. If the intra-cluster gas density is high enough, e.g. near the centre of a cluster, \citet{QuiMooBow00} conclude that ram pressure stripping can remove 100\% of the atomic gas content in massive galaxies in a time scale as fast as $10^8$ yr. The hot intra-cluster gas may also remove any available gaseous halo around the galaxy that would otherwise fall into the galaxy and feed further star formation \citep[see][]{LarTinCal80,BekCouShi02}. Such process, dubbed starvation or strangulation, is able to destroy such a gaseous halo even away from the cluster core, although in this case on a time scale of $\approx1$ Gyr. Another effect, dubbed harassment, can also remove an existing gaseous halo via the weak interaction with relatively nearby galaxies in e.g. a dense cluster core \citep[e.g.][]{Ric76,MooLakKat98}. The time scales for harassment to be effective in changing the stellar population content in a disc galaxy are of the order of a few Giga-years. In summary, given the position of the galaxy in the cluster, it is not unlikely that NGC 4371 has undergone some ram pressure stripping and starvation, which would explain the mean old ages of its stars \citep[see e.g.][]{VolCayBal01,KenGehJac14}.

Whatever the mechanism that made the inner $1'$ squared of NGC 4371 devoid of fuel to produce new stars, it was completed after the formation of the bar and the subsequent gas inflow to form the inner disc and the $10''$ ring. There is no need to invoke ram pressure stripping to cease star formation in the area within the bar radius, however. The bar is able to push gas to the centre fast \citep[at a typical time scale of $\sim10^8$ yr,][see also \citealt{HuaKau15,GavConDot15}]{Ath92,EmsRenBou15}, and the subsequent formation of the inner disc and nuclear ring and other nuclear disc-like structures might very well be able to consume all available gas. The bar also restricts the flow of gas coming from outside its radius in the plane of the disc \citep[see][]{BouCom02}. However, it is necessary to prevent that gas external to the galaxy falls in the disc within the bar radius from a direction that is not aligned with the disc plane. Starvation/strangulation and harassment come naturally to mind.

The youngest spatial bins in the disc have a mean stellar age of about 7 Gyr (see Fig. \ref{fig:distr}), corresponding to a formation redshift of $z$ about 0.8. This would mean that the intra-cluster gas was already exerting pressure at such early times, when the formation of the cluster was still at an early stage. One should also consider the likely possibility that strong environmental processes could have already occurred if NGC 4371 was part of a group before entering the cluster \citep[see][]{HaiPerSmi15}. In any case, such environmental effects must have been completed at a redshift of about 0.8.

Since the bar has some spatial bins with mean stellar age as low as 8 Gyr (corresponding to a formation redshift of about unity, Fig. \ref{fig:distr}), remnant gas promoted some star formation in the bar itself after the formation of the inner disc and the $10''$ ring, but before the completed quenching at a redshift of about 0.8. In this context, it is also interesting to note that the difference in mean stellar age between the peak of the distribution for the $10''$ ring and the younger spatial bins present there is about 1 Gyr. One expects that the gas brought to the centre by the bar gets compressed at its ILR and form stars, forming the ring, but also originating supernova events that will put some gas back in the inter-stellar medium. This may result in a second generation of stars, which is probably manifested in the lower age tail of the mean stellar age distribution of the $10''$ ring. This newest generation of stars could be the reason why the mean stellar age of the $10''$ ring is slightly less than that of the inner disc (see Fig. \ref{fig:struct_pops}). Furthermore, this second generation of stars may also be the reason why the mean metallicity of the $10''$ ring is slightly elevated, as compared to that of the inner disc (again, Fig. \ref{fig:struct_pops}).

Interaction with the intra-cluster medium thus appears to have played a major role in the evolution of NGC 4371, quenching star formation. From a structural perspective, however, the galaxy does not present signs of interactions with other galaxies or mergers \citep[but see][who suggested that disc-dominated galaxies without classical bulges can be formed via major mergers]{WanHamPue15}. In fact, the evolution of the galaxy seems markedly governed by internal processes, which is intriguing considering its position at the core of the Virgo cluster. One possibility is that the galaxy is coming near the core of the cluster for the first time, but to test this is beyond the scope of this study. NGC 4371 is, nevertheless, an example of a galaxy whose evolution is strongly influenced by internal processes despite residing in a dense environment.

\section{Summary and conclusions}
\label{sec:conc}

We have used VLT/MUSE integral field spectroscopy data, combined with imaging data from the HST/ACS and Spitzer/IRAC imaging cameras, in order to investigate the kinematics, stellar population content and structural properties of NGC 4371, an early-type massive barred galaxy in the core of the Virgo cluster. The MUSE spectral coverage and resolution, allied to its fine spatial resolution, allows us to make robust statements about the stellar population content in the inner $1'$ squared of the galaxy. We summarise our main conclusions as follows:

\begin{enumerate}
\item We find that the mean stellar age in the inner disc and $10''$ ring close to the centre of the galaxy is above 10 Gyr, with a scattered and negligible contribution from younger stars. With the safe assumption that the stars in this ring were formed by gas brought to the centre  -- via bar-induced secular evolution processes -- and compressed at or near the bar ILR, this result indicates that the bar in NGC 4371 was formed at a redshift $z=1.8^{+0.5}_{-0.4}$, and has been ever since influencing the evolution of the galaxy.
\item We find that the mean stellar age of the more central part of the galaxy major disc (our MUSE field does not cover the whole galaxy) is above 7 Gyr, with a small contribution of younger stars. We interpret this result in the context of the evolutionary processes connected to the high-density environments in the Virgo cluster. This result thus suggests that the removal of gas from the galaxy (ram-pressure stripping) and the removal of external cold gas available to fall onto the disc and ignite more star formation (starvation/strangulation) were already effective at a redshift  $z=0.8^{+0.2}_{-0.1}$, even though the cluster is still not a dynamically relaxed system today, at $z=0$. The error bars are derived from the Monte Carlo realisations described at the end of Sect. \ref{sec:ages1}.
\end{enumerate}

With these data alone, we cannot say much about the fraction of barred galaxies at $z>1$. However, our results on NGC 4371 do open up the possibility that -- even if bar-driven secular evolution was not the dominant galaxy evolution process at these redshifts, being overshadowed by galaxy interactions -- bar-driven secular evolution may have a more extended and significant role in the history of disc galaxies, not necessarily restricted to more recent cosmic epochs.

It is evident the necessity of pursuing a similar study as the current one, but strengthened with a large, statistically significant sample of nearby disc galaxies. Furthermore, in the case of NGC 4371 for example, mapping the galaxy to expand the area observed with MUSE is potentially very revealing. With such a mosaic of MUSE data cubes, the tentative detection of the h$_4$ minima along the bar, discussed in Sect. \ref{sec:kin2}, could be put on firmer grounds, since they seem to occur just at the outskirts of our MUSE field. In addition, this work also signals the edge that will be gained once the narrow field mode in MUSE is ready, providing a spatial resolution of the order of $0.05''$. At the distance of NGC 4371, this corresponds to an astonishing spatial resolution of only 4 pc, which will clearly unveil the nature of the innermost stellar components in the galaxy, and their connected evolution with the central supermassive black hole.

\begin{acknowledgements}
This work would have been impossible without the vigorous work of Fernando Selman and George Hau, MUSE Paranal instrument scientists. We are grateful to Francesca Iannuzzi and Francesca Fragkoudi for very useful input and discussions on bars in simulated galaxies, particularly box/peanuts. It is also a pleasure to thank Eric Emsellem for useful discussions on the inner components of barred galaxies, and Michael Vlasov for his permission to use a chart from his atlas. We are very grateful to the anonymous referee for a thoughtful and constructive report. M.K.S. and J.F-B acknowledge support from grant AYA2013-48226-C3-1-P from the Spanish Ministry of Economy and Competitiveness (MINECO). J.F-B also acknowledges support from the DAGAL network from the People Programme (Marie Curie Actions) of the European Union's Seventh Framework Programme FP7/2007-2013/ under REA grant agreement number PITN-GA-2011-289313. P.S-B acknowledges support from the Ram\'on y Cajal program, grant ATA2010-21322-C03-02 from the Spanish Ministry of Economy and Competitiveness (MINECO). Based on observations made with ESO Telescopes at the La Silla Paranal Observatory under programme ID 060.A-9313.
\end{acknowledgements}

\bibliographystyle{aa} % style aa.bst
\bibliography{../../../../../gadotti_refs} % your references Yourfile.bib

\end{document}